\documentclass[iop,twocolappendix]{emulateapj}
\usepackage{blindtext}
\usepackage{placeins}
\usepackage{graphicx}
\usepackage{amssymb, amsmath}
\usepackage{natbib}
\usepackage{amsmath}
\usepackage{bm}
\usepackage{color}
\usepackage[colorlinks,urlcolor=blue,citecolor=blue,linkcolor=blue]{hyperref}
\usepackage[caption=false]{subfig}

\begin{document}
\title{Electron and Proton Acceleration in Trans-Relativistic Magnetic Reconnection: Dependence on Plasma Beta and Magnetization}

\author{David Ball,\altaffilmark{1,3} Lorenzo Sironi,\altaffilmark{2} and Feryal \"Ozel\altaffilmark{1}}

\altaffiltext{1}{Department of Astronomy and Steward Observatory, Univ. of Arizona, 933 N. Cherry Avenue, Tucson, AZ 85721, USA}

\altaffiltext{2}{Department of Astronomy, Columbia University, 550 West 120th Street, New York, NY 10027, USA}
\altaffiltext{3}{Email: davidrball@email.arizona.edu}

\begin{abstract}
Non-thermal electron acceleration via magnetic reconnection is thought to play an important role in powering the variable X-ray emission from radiatively inefficient accretion flows around black holes, such as Sgr A* at our Galactic center.  The trans-relativistic regime of magnetic reconnection, where the magnetization $\sigma$, defined as the ratio of magnetic energy density to enthalpy density, is $\sim 1$, is frequently encountered in such flows.
By means of a large suite of two-dimensional particle-in-cell simulations, we investigate electron and proton acceleration in the trans-relativistic regime.  We focus on the dependence of the electron energy spectrum on $\sigma$ and the proton $\beta$ (i.e., the ratio of proton thermal pressure to magnetic pressure). We find that the electron spectrum in the reconnection region is non-thermal and can be generally modeled as a power law.  At $\beta \lesssim 3 \times 10^{-3}$, the slope, $p$, is independent of $\beta$ and it hardens with increasing $\sigma$ as $p\simeq 1.8 +0.7/\sqrt{\sigma}$. Electrons are primarily accelerated by the non-ideal electric field at X-points, either  in the initial current layer  or in current sheets generated in between merging magnetic islands. 
At higher values of $\beta$, the electron power law steepens and the electron spectrum eventually approaches a Maxwellian distribution for all values of $\sigma$. At values of $\beta$ near $\beta_{\rm max}\approx1/4\sigma$, when both electrons and protons are relativistically hot prior to reconnection, the spectra of both species display an additional component at high energies, containing a few percent of particles.  These particles are accelerated via a Fermi-like process by bouncing in between the reconnection outflow and a stationary magnetic island.
For the main population of non-thermal electrons that excludes this additional component,
we provide an empirical prescription for the dependence of the power-law slope and the acceleration efficiency on $\beta$ and $\sigma$, which can be used in global simulations of collisionless accretion disks.  
\end{abstract}
\maketitle

\section{Introduction}
Magnetic reconnection is widely thought to play an important role in the episodic flaring activity of numerous astrophysical systems, including blazar jets (\citealt{giannios2013}; \citealt{petropoulou2016}; \citealt{nalewajko2016}), pulsar wind nebulae (\citealt{coroniti1990}; \citealt{lyubarsky2001}; \citealt{zenitani2001}; \citealt{kirk2003}; \citealt{contopoulos2007}; \citealt{petri2008};  \citealt{sironi2011}; \citealt{cerutti2012, cerutti2014, cerutti2017}; \citealt{philippov2014}; see \citealt{sironi2017} for a recent review), gamma ray bursts (\citealt{thompson1994, thompson2006}; \citealt{usov1994}; \citealt{spruit2001}; \citealt{drenkhahn2002}; \citealt{lyutikov2003}; \citealt{giannios2008}), the Sun (\citealt{forbes1996}; \citealt{yokoyama2001}; \citealt{shibata2011}), and accretion flows around black holes (\citealt{galeev1979}; \citealt{dimatteo1998}; \citealt{uzdensky2008}; \citealt{li2015}; \citealt{ball2016}; \citealt{li2017}).  Through magnetic reconnection, energy stored in magnetic fields is able to dissipate into the ambient plasma, resulting in particle heating and acceleration. Electrons accelerated to ultra-relativistic energies can produce flares and high-energy emission.  Many of these astrophysical systems consist of low-density ``collisionless'' plasmas, where the timescale for Coulomb collisions is significantly longer than dynamical timescales. Here, the dynamics and energetics of magnetic reconnection can  be properly captured only by means of a fully-kinetic framework, which can be achieved via numerical techniques such as particle-in-cell (PIC) simulations.

One of the key parameters that determines the outcome of reconnection and the properties of the resulting particle distribution is the  magnetization of the ambient plasma, i.e., the ratio $\sigma$ of magnetic energy density to enthalpy density. 
Numerous studies have investigated the non-relativistic ($\sigma\ll 1$) regime, which has applications to the solar corona and solar flares (e.g., \citealt{drake2013}; \citealt{dahlin2014}; \citealt{shay2014}; \citealt{liguo2015}).  The ultra-relativistic ($\sigma\gg 1$) regime has also been explored in detail, due to its relevance to high-energy emission from blazar jets and pulsar wind nebulae (e.g., \citealt{kagan2013}; \citealt{sironi2014}; \citealt{guod_14, guo2015, guo_16}; \citealt{melzani2014}; \citealt{kagan2015}; \citealt{nalewajko2015}; \citealt{werner2016}; \citealt{sironi2016}).  However, only a limited number of studies have been carried out in the trans-relativistic regime ($\sigma \sim 1$), addressing particle heating (\citealt{rowan2017}) and acceleration (\citealt{melzani2014b}; \citealt{werner2018}). The trans-relativistic regime is of particular interest to studies of radiatively inefficient accretion flows around black holes, such as Sgr A* at our Galactic center. Here, current sheets with typical magnetizations of $\sigma\sim1$ are frequently observed in global MHD simulations (\citealt{ball2017}).  Localized particle acceleration powered by magnetic reconnection in these settings could give rise to high-energy variability (\citealt{ball2016}; \citealt{li2017}).  

Earlier investigations (e.g., \citealt{schoeffler2011, schoeffler2013}; \citealt{rowan2017}) have shown that, in addition to the magnetization $\sigma$, the initial plasma temperature, or equivalently the proton $\beta$ (i.e., the ratio of proton thermal pressure to magnetic pressure), can affect the dynamics and energetics of magnetic reconnection.  In particular, \citet{rowan2017} explored the dependence of the electron and proton heating efficiency on the magnetization $\sigma$, the proton $\beta$, and the electron-to-proton temperature ratio.  However, the role of $\beta$ on non-thermal particle acceleration in the trans-relativistic regime ($\sigma\sim1$) remains largely unexplored. 

The works by \citet{melzani2014b, melzani2014} were the first to investigate particle acceleration in the trans-relativistic regime of reconnection. They examined the energy partition between protons and electrons and the electron power-law spectra, but they employed a reduced proton-to-electron mass ratio and they only explored a relatively narrow range of $\beta$. \citet{werner2018} performed an extensive study across a wide range of $\sigma$, from the trans-relativistic through the ultra-relativistic regime, and reported how the reconnection rate, the electron power-law slope, and the energy partition between electrons and protons depend on $\sigma$.  They found that the electron power-law slope decreases with increasing $\sigma$ (i.e., the spectrum hardens) and provided an empirical fit $p\simeq1.9+0.7/\sqrt{\sigma}$ for the power-law slope $p$ of the electron spectrum.  However, their study was performed at a fixed value of proton beta $\beta=0.01$.  


In this work, we investigate proton and electron non-thermal acceleration in trans-relativistic  reconnection, covering the whole parameter space in $\sigma$ and $\beta$ and employing the physical proton-to-electron mass ratio. For four values of the magnetization ($\sigma=$ 0.1, 0.3, 1, and 3), we explore a wide range of $\beta$, from $\beta=10^{-4}$ up to the maximum possible value of $\beta$, that is $\beta_{\rm max}\approx 1/4\sigma$ (we will discuss why this is the case in Section \ref{setup}). Our study goes beyond the current state of the art in several respects: we explore for the first time the dependence of the non-thermal electron spectrum on  plasma $\beta$ and we examine the role of various electron acceleration mechanisms, by tracking particles in our simulations. In addition, our computational domains are larger than previous works by at least a factor of $\sim5$.
While we primarily focus on electrons, we also present proton spectra and briefly investigate proton acceleration mechanisms.

We find that the electron spectrum in the reconnection region can be generally modeled as a non-thermal power law, but the properties of the spectrum are strongly dependent on $\beta$. At $\beta \lesssim 3 \times 10^{-3}$, the spectrum is dominated by a hard power law, whose slope is insensitive to $\beta$ and depends on $\sigma$ as $p\simeq 1.8 +0.7/\sqrt{\sigma}$, in agreement with the result by \citet{werner2018}.  Electrons are primarily accelerated by the non-ideal electric field at X-points, either  in the initial current layer  or in current sheets generated in between merging magnetic islands. 
At higher $\beta$, the electron power law steepens significantly, and the electron spectrum gradually approaches a Maxwellian distribution, for all values of $\sigma$. At the highest values of $\beta$ near $\beta_{\rm max}\approx1/4\sigma$, when both electrons and protons start relativistically hot, the spectrum of both species displays an additional component at high energies, containing a few percent of particles, which are accelerated via a Fermi-like process by bouncing in between the reconnection outflow and the stationary magnetic island at the boundary of our periodic domain.
For the main population of non-thermal electrons (i.e., excluding the additional component emerging at $\beta\rightarrow \beta_{\rm max}$),
we provide an empirical prescription for the dependence of the power-law slope and the acceleration efficiency on $\beta$ and $\sigma$. The results of our study can be used as subgrid models in global MHD simulations of  black hole accretion flows (e.g., \citealt{ball2016}; \citealt{mao2017}; \citealt{chael2017}), potentially unveiling the origin of the flaring behaviour of Sgr A* \citep{ponti2017}.

We also investigate the dependence of our results on the size of the simulation domain, and find that the high-energy cutoff of the electron spectrum increases with box size.  Additionally, we find that the electron spectra tend to steepen for larger simulation domains, with a relatively weak dependence of the power-law slope on box size.

The layout of the paper is as follows.  In Section \ref{setup} we describe the setup of our simulations.  In Section \ref{time_evol}, we show and describe the time evolution of a representative simulation and the evolution of its associated electron and proton energy spectra.  In Section \ref{role_of_beta}, we explore the dynamics of the reconnection layer as a function of $\sigma$ and $\beta$, and illustrate the key differences between low- and high-$\beta$ reconnection.  In Section \ref{energy_spectra}, we show the electron and proton spectra for a number of values of $\sigma$ and $\beta$, and provide an empirical fit to the electron power-law slopes and acceleration efficiencies.  Finally, in Section \ref{mechanism}, we show representative trajectories of accelerated electrons for both a low-$\beta$ and a high-$\beta$ simulation.  We conclude and summarize in Section \ref{conclusions}.

\section{Simulation Setup}\label{setup}
We perform a large suite of PIC simulations of anti-parallel magnetic reconnection using the  publicly-available code TRISTAN-MP (\citealt{buneman93}; \citealt{spitkovsky05}).  We employ a two-dimensional (2D) simulation domain in the $xy$ plane, but we track all three components of velocity and electromagnetic field vectors.  We set up the system in Harris equilibrium, with a magnetic field profile $\bm{{B}} = -B_{0}\tanh{\left(2\pi y / \Delta\right)}\bm{\hat{x}}$, where $B_{0}$ is the strength of the reconnecting field in the ambient plasma and $\Delta$ is the thickness of the sheet.  $B_{0}$ is related to the magnetization parameter $\sigma$ via $\sigma=B_{0}^2 / 4\pi w_{0} $, where $w_{0}$ is the enthalpy density of the ambient plasma $w_0=(\rho_{e}+\rho_{i})c^{2}+ \hat{\gamma}_{e}u_{e}+ \hat{\gamma}_{i}u_{i}$, with $\rho_{i,e}$, $\hat{\gamma}_{i,e}$, and $u_{i,e}$ being the mass densities, adiabatic indices, and internal energy densities of ambient protons and electrons, respectively.  The temperature is specified through the proton $\beta$, defined as $\beta \equiv \beta_{i}=8 \pi n_{i} k T_{i}/B_{0}^{2}$, where $n_i=\rho_i/m_i$ is the proton number density, $T_i$ is the proton temperature, and $m_i$ is the proton mass. Ambient electrons and protons start with the same temperature, such that $\beta_e=\beta_i=\beta$ (the total plasma-$\beta$, including both species, is $2\,\beta$). In most cases, the ambient protons are non-relativistic, so the magnetization parameter as defined with the proton rest mass $\sigma_i=B_0^2/4\pi \rho_i c^2$ is nearly identical to the enthalpy-weighted magnetization $\sigma$ defined above. Each computational cell in the ambient plasma is initialized with four particles per cell (i.e., $N_{ppc}=4$), but we have tested that our results are the same when using $N_{ppc}=16$ (see Appendix \ref{convergence}).

The thickness of the current sheet is $\Delta = 80 \,c/\omega_{p}$, where $\omega_{p}$ is the electron plasma frequency, given by 

\begin{equation}
\omega_{p} =\sqrt{\frac{4\pi n_{e}e^{2}}{m_{e}}}\left(1+\frac{\theta_{e}}{\hat{\gamma_{e}}-1}\right)^{-1/2}.
\label{electron skin depth}
\end{equation}
Here, $e$ is the electron charge, $m_e$ is the electron mass, $n_e=\rho_e/m_e$ is the electron number density ($which is also equal to the proton number density n_i$) and $\theta_{e}$ is the dimensionless electron temperature $\theta_{e}=kT_{e}/m_{e}c^{2}$ in the ambient plasma. 
We set up the initial Harris equilibrium by initializing the plasma in the current sheet to be hot and overdense (by a factor of 3 with respect to the background) so that its thermal pressure balances the magnetic pressure outside the sheet. The particles in the current sheet are set up as a Maxwellian distribution drifting in the $z$ direction, so that their electric current balances the curl of the magnetic field.  The hot particles that we set up in the current layer are never included in the particle energy spectra that we present below, since their properties depend on our specific choice of initialization of the Harris sheet.

Our computational domain is periodic in the $x$-direction of the reconnection outflow in order to retain all accelerated particles, while the box is continually enlarged in the $y$-direction, as two moving injectors --- that steadily inject magnetized plasma into the simulation domain --- recede from the current sheet at the speed of light along $\pm\bm{\hat{y}}$. By employing the moving injectors and a dynamically-enlarging box (see \citealt{sironi2009} for further details), we can study the late-time evolution of the system without being artificially limited by the finite amount of plasma and magnetic flux that is initially in the simulation domain. Additional computational optimization is achieved by allowing the injectors to periodically ``jump'' back towards the current sheet, removing all particles beyond the injectors and resetting the electromagnetic fields to their initial values.

The length of the box in the $x$ direction of the reconnection outflow is $L_x=16620$ cells, which corresponds to $L_x\simeq5540\,c/\omega_{p}$ because we resolve the electron skin depth $c/\omega_{p}$ with 3 computational cells. As we describe in Appendix \ref{convergence}, we have tested that our results are the same when the electron skin depth is resolved with 6 cells. We also investigate the dependence of our results on the extent $L_x$ of the computational domain (up to a factor of two larger than our reference runs; see Appendix \ref{boxsize}).

We trigger reconnection in the center of the box by removing instantaneously the pressure of the hot particles initialized in the center of the sheet (see \citealt{sironi2016}).  This causes the current sheet to collapse and form two ``reconnection fronts,'' which are pulled by magnetic tension along $\pm \bm{\hat{x}}$ at roughly the Alfv\'en speed $v_{A}=c\sqrt{\sigma/(1+\sigma)}$.  We define the Alfv\'enic crossing time as  $t_{A} = L_{x}/v_{A}$.  At $t\gtrsim 0.5\,t_A$, the reconnected plasma starts accumulating at the boundary of the periodic simulation domain, where a ``boundary island'' forms.

Astrophysical current sheets are likely to be thick, making the timescale for spontaneous (or ``untriggered'') reconnection very long compared to relevant dynamical timescales.  This means that astrophysical reconnection is likely triggered by some large-scale perturbation, which motivates our decision to trigger reconnection in our simulations. In fact, the large-scale perturbation will induce a curvature of the field lines over a scale $\sim L_x$, such that the current sheet is narrower near the center. The central region is then most likely to go unstable via the tearing mode, and the signal of ongoing reconnection will propagate toward the outer regions (where the current sheet is broader) before they have time to spontaneously become unstable. We further discuss our choice of a triggered setup in  Appendix \ref{untriggered}, where we compare our results to the case of untriggered reconnection, where the  system goes unstable via numerical noise, as in \citealt{sironi2014}. In Appendix \ref{untriggered}, we also compare the results of triggered simulations with either periodic or outflow boundaries in the $x$ direction (for further details on the implementation of the outflow boundary conditions, see \citealt{sironi2016}).

\begin{deluxetable*}{ccccccc}
\centering
\tablewidth{1.5\columnwidth}
\tablecaption{Simulation Parameters}  
\tablehead{run&$\sigma$& $\sigma_{i}$ & $\beta$ & $L_{x}$ ($c/\omega_{pi}$) &$L_{x}$ ($r_{e,\rm hot}$) &$kT_{i}/m_{i}c^{2}$}
\startdata
A0& 0.1 & 0.1& $1\times 10^{-4}$ & 125 & 406&$5 \times 10^{-6}$\\ 
 \hline
A1& 0.1 & 0.1& $3\times 10^{-4}$ & 127 & 417&$1.5 \times 10^{-5}$\\ 
 \hline
A2& 0.1& 0.1 & $10^{-3}$ & 134& 453 & $5 \times 10^{-5}$\\
 \hline
A3& 0.1& 0.1 & $3\times 10^{-3}$ & 158 & 542 & $1.5 \times 10^{-4}$\\
 \hline
A4& 0.1& 0.1 & 0.01  & 233& 776 & $5 \times 10^{-4}$\\
 \hline
A5&  0.1 & 0.1& 0.02& 312 & 1020& $1 \times 10^{-3}$\\
 \hline
A6&  0.1 &  0.1 &0.1 & 664& 2110 & $5 \times 10^{-3}$\\
 \hline
A7 & 0.1& 0.11 & 0.3 & 1138 & 3978& 0.02\\
 \hline
A8& 0.1 &0.16& 1.5  & 4133& 7269 & 0.1\\ 
 \hline
B0 & 0.3&0.3 & $1\times 10^{-4}$ & 127& 241&$1.5\times 10^{-5}$\\ 
 \hline
B1 & 0.3&0.3 & $3\times 10^{-4}$  & 134& 261&$5\times 10^{-5}$\\ 
 \hline
B2& 0.3&0.3 & $10^{-3}$  & 156& 313&$1.5\times 10^{-4}$\\
  \hline
B3& 0.3&0.3 & $3\times 10^{-3}$ & 232& 448&$5\times 10^{-4}$\\
 \hline
B4& 0.3&0.3 & $6\times 10^{-3}$ &312& 589 &$1\times 10^{-3}$\\
 \hline
B5& 0.3&0.3 & 0.01 &375& 701&$1.5\times 10^{-3}$\\
 \hline
B6&  0.3&0.3 & 0.03 & 664& 1218&$5\times 10^{-3}$\\
 \hline
B7&  0.3 & 0.34&0.11 & 1138& 2296&0.02\\
 \hline
B8*& 0.3&0.72 & 0.55 & 4133& 4956&0.2\\ 
 \hline
C0&  1 &1& $1\times 10^{-4}$ & 134& 143&$5 \times 10^{-5}$\\ 
 \hline
C1&  1 &1& $3\times 10^{-4}$ & 157& 171&$1.5 \times 10^{-4}$\\ 
 \hline
C2& 1 &1& $10^{-3}$  &232& 245&$5 \times 10^{-4}$\\
 \hline
C3& 1 &1& $3\times 10^{-3}$ & 375& 384&$1.5 \times 10^{-3}$\\
 \hline
C4& 1 &1& 0.01 & 664& 667&$5 \times 10^{-3}$\\
 \hline
C5& 1&1.1 & 0.03 & 1138& 1107& 0.015\\
 \hline
C6& 1&1.3 & 0.08  & 2069& 1827& 0.05\\
 \hline
C7*& 1&2.4 & 0.16 & 4133& 2713&0.2\\
  \hline
D0&  3&3 & $10^{-4}$ & 157& 99 & $1.5 \times 10^{-4}$\\ 
 \hline
D1& 3 &3& $3\times 10^{-4}$ & 232& 141& $5 \times 10^{-4}$\\ 
 \hline
D2& 3 &3& $10^{-3}$ & 375& 221& $1.5 \times 10^{-3}$\\
 \hline
D3& 3 &3.1& $3\times 10^{-3}$ & 664& 385& $5 \times 10^{-3}$\\
 \hline
D4& 3&3.3 & 0.01 & 1138& 639&0.015\\
 \hline
D5*&  3&4.0 & 0.026 & 2069& 1055&0.05\\
 \hline
D6*&  3 &7.2 &0.055  & 4133& 1566&0.2
\enddata
\tablecomments{Summary of the physical and numerical parameters of our simulations.  All simulations are performed with the physical mass ratio, equal electron and proton temperatures, a resolution of 3 cells per electron skin depth, and 5,440 electron skin depths along the current layer.}
\label{tab:fit}
\end{deluxetable*}

The physical and numerical parameters of our simulations are summarized in Table 1. To fully map out the parameter space of interest, we perform 33 simulations spanning four different values of the magnetization: $\sigma=\,$0.1, 0.3, 1, and 3. For each value of $\sigma$, we have multiple (at least 7) simulations in which we vary the proton $\beta$ from $10^{-4}$ up to the maximum possible value of $\beta$, $\beta_{\rm max}\approx 1/4\sigma$. This upper limit in $\beta$ is reached when both protons and electrons become relativistically hot. In this limit, the internal energy densities dominate over the rest mass energy densities, such that the enthalpy density can be written as  $w_0\simeq \hat{\gamma}_{e}u_{e}+\hat{\gamma_{i}}u_{i}$. For $\hat{\gamma}_{e}=\hat{\gamma}_{i}=4/3$, as appropriate for a 3D ultra-relativistic gas, the magnetization tends to $\sigma \simeq 1/4\beta$, which defines an upper limit on $\beta$ at a given $\sigma$, equal to $\beta_{\rm max}\approx 1/4\sigma$.

Due to computational constraints, PIC codes often employ a reduced proton-to-electron mass ratio, in order to decrease the separation of scales between the two species. However, as shown in \citealt{rowan2017}, a choice of the mass ratio smaller than the physical one can artificially affect the partition of energy between electrons and protons in trans-relativistic reconnection. Since it could also artificially affect the efficiency and slope of non-thermal particle acceleration, we employ the physical mass ratio $m_{i}/m_{e}=1836$ in this study. While the box length $L_x$ measured in electron skin depths is independent of $\beta$ or $\sigma$, the box length in proton skin depths $c/\omega_{pi}$, where the proton plasma frequency is given by
$$
\omega_{pi} =\sqrt{\frac{4\pi n_{i}e^{2}}{m_{i}}}\left(1+\frac{\theta_{i}}{\hat{\gamma_{i}}-1}\right)^{-1/2},
$$ 
varies significantly with $\beta$ due to the $\theta_e$-dependent correction in Equation \ref{electron skin depth}.  For most of our simulations, electrons start as ultra-relativistically hot, while protons are non-relativistic (our maximum $\theta_{i}=kT_i/m_i c^2$ is 0.2, see Table 1).  As  $\beta$ increases, the separation of electron and proton scales decreases, so our domain is effectively larger in units of the proton skin depth at higher $\beta$ (see Table 1). In Table 1, we also quote the extent of our simulation domain in units of $r_{e,\rm hot}=\sigma_e m_e c^2/eB_0$ (the unit of length employed by \citealt{werner2018}), which corresponds to the Larmor radius of a relativistic electron with Lorentz factor $\sigma_e=\sigma_i m_i/m_e$.

\begin{figure}[!h]
\includegraphics[width=1.0\linewidth]{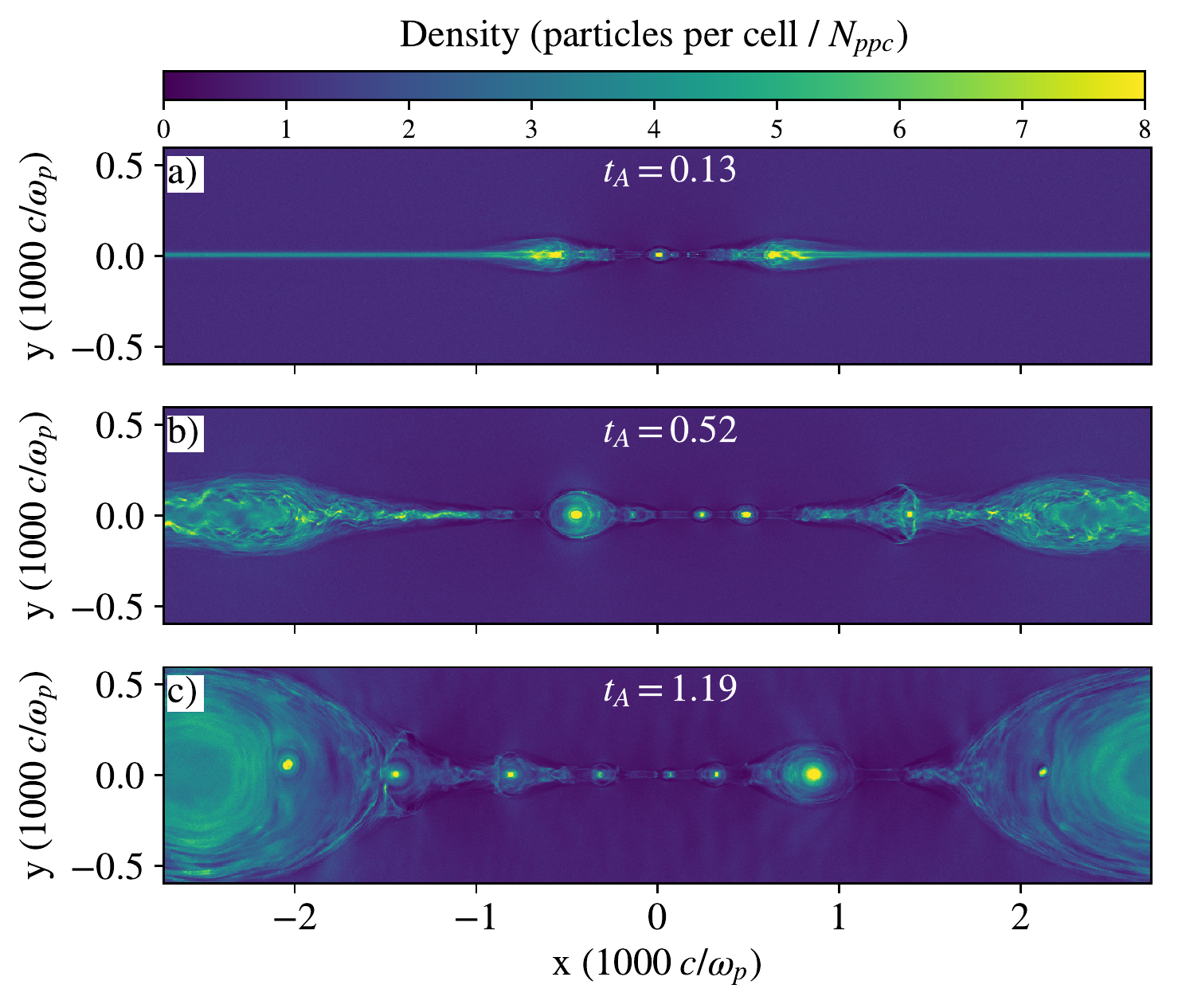}
\caption{2D snapshots of density depicting the time evolution from a simulation with $\sigma=0.3$ and $\beta=3\times 10^{-4}$ (run B1) at three different times.  The top, middle, and bottom panels correspond to 0.13, 0.52, and 1.19 Alfv\'en crossing times, respectively.  We normalize the density to the initial number of particles per cell in the ambient plasma, $N_{ppc}$. }
\label{lowbeta_threeplot}
\end{figure}

\section{Time Evolution of the reconnection layer}\label{time_evol}
In order to illustrate the time evolution of a typical simulation, we show in Figure \ref{lowbeta_threeplot} a series of 2D snapshots of the particle number density for a run with $\sigma=0.3$ and $\beta=3\times10^{-4}$. The lack of pressure support in the vicinity of the center, resulting from our initial perturbation, triggers the collapse of the current sheet (top panel in Figure \ref{lowbeta_threeplot}) and the formation of an X-point. In the following, we shall indicate this X-point as the ``primary X-point.'' While in untriggered systems the tearing mode instability pinches the current sheet at several locations, thus producing several primary X-points, in our triggered setup we only have one primary X-point. The top panel also shows the two reconnection fronts, at $x \approx \pm 500\, c/\omega_{p}$, that are pulled towards the edges of the box by the tension of the magnetic field lines.  In the underdense region in between these fronts, a secondary plasmoid begins to form close to the center, as plasma flows in from above and below the  reconnection layer.  

In the middle panel of Figure \ref{lowbeta_threeplot}, the reconnection fronts are approaching the edges of the box and numerous secondary plasmoids have formed in the layer, separated by secondary X-points (see e.g., \citealt{loureiro2007}; \citealt{uzdensky2010}; \citealt{huang2012}; \citealt{takamoto2013} and \citealt{comisso2016} for the physics of secondary plasmoid formation).  By ``secondary plasmoids,'' we refer to structures that form after the early collapse of the current sheet and that contain particles that belong to the ambient plasma (as opposed to the hot population of particles initialized in the current sheet). A secondary X-point is present in between each pair of neighboring secondary plasmoids. 
The largest plasmoid near the center of the box, at $x\simeq-300 \; c/\omega_{p}$, is formed via mergers of several smaller plasmoids, and contains the highest energy particles in the system at this time (see \citealt{sironi2016} for a discussion of the correlation between plasmoid size and  maximum particle energy).  

In the final snapshot (bottom panel in Figure \ref{lowbeta_threeplot}), the outflowing fronts collide across the periodic boundaries, forming a large magnetic island that sits passively at the edge and acts as a reservoir for accelerated particles. In the following, we shall refer to this structure as the ``boundary island,'' and reserve the term``plasmoids'' to refer only to secondary islands. Secondary plasmoids forming in the sheet are eventually advected into the boundary island by the  tension of the field lines.   A current sheet forms at the interface between the boundary island and each secondary plasmoid that is merging into it. As we will show in Section \ref{mechanism}, this interface is a site of efficient electron acceleration.

\subsection{Defining the Reconnection Region}
\begin{figure}[!h]
\includegraphics[width=\linewidth]{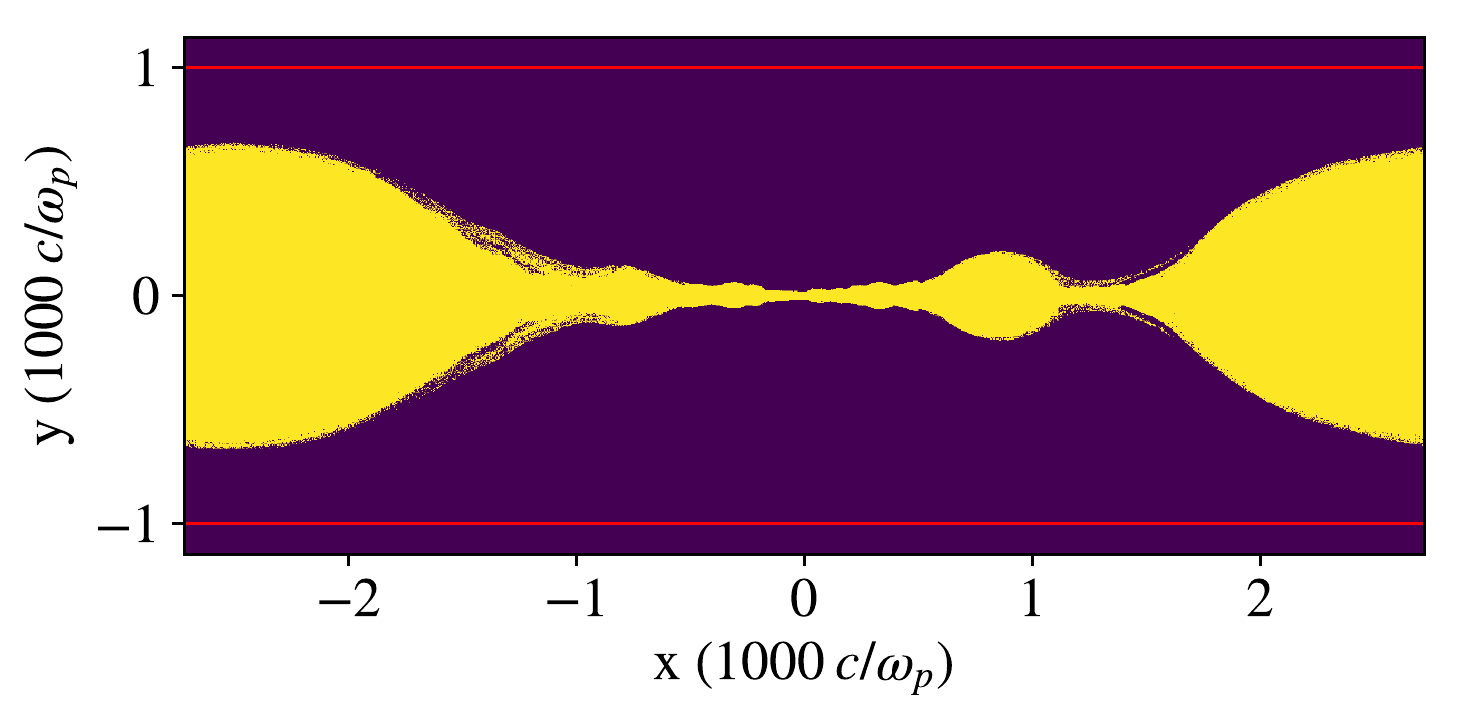}
\caption{Reconnection region (yellow) identified based on the mixing criterion described in the text. The snapshot refers to the same simulation shown in Figure \ref{lowbeta_threeplot} ($\sigma=0.3$, $\beta=3\times 10^{-4}$, B1), at the last time shown (bottom panel in Figure \ref{lowbeta_threeplot}).  We see that our mixing criterion properly isolates the reconnected overdense plasma from the cold upstream plasma.  The red lines delimit the region in which the total spectra are calculated, as we will discuss in Figure~\ref{timespec}.}
\label{regions}
\end{figure}

The spectrum from the entire simulation domain includes both pre- and post-reconnection plasma.  Because of this, it is prudent to have a scheme to distinguish between particles that have undergone reconnection, and particles that are still in the colder upstream region.  This is an important step to correctly interpret the spectra and avoid mistaking for a power-law component the ``bridge'' between the pre- and post-reconnection distributions.
In order to extract the spectrum of the plasma that has undergone reconnection, uncontaminated by the cold upstream plasma, we use a mixing criterion to identify regions where reconnection has occurred (as first proposed in \citealt{daughton2014}, and described in \citealt{rowan2017}).  In short, we tag particles with an identifier that specifies whether they were initialized above or below the current sheet.  We can then identify the cells where particles have mixed to a sufficient degree and in doing so, define the ``reconnection region'', predominantly populated by particles that have been processed by reconnection.   We take a mixing fraction of one part in 100 as our lower limit to define this region.  Using this technique, we are able to cleanly separate the particles that are part of a region that has undergone reconnection from the colder upstream plasma. For the remainder of this paper, any reference to the ``reconnection region'' refers to the region defined by this criterion.  We show in Figure \ref{regions} the result of applying this method to the snapshot shown in the bottom panel of Figure \ref{lowbeta_threeplot}.  We see that the reconnection region (yellow) is cleanly separated from the upstream plasma (dark purple).

\begin{figure}[htp]
{
	\includegraphics[width=\linewidth]{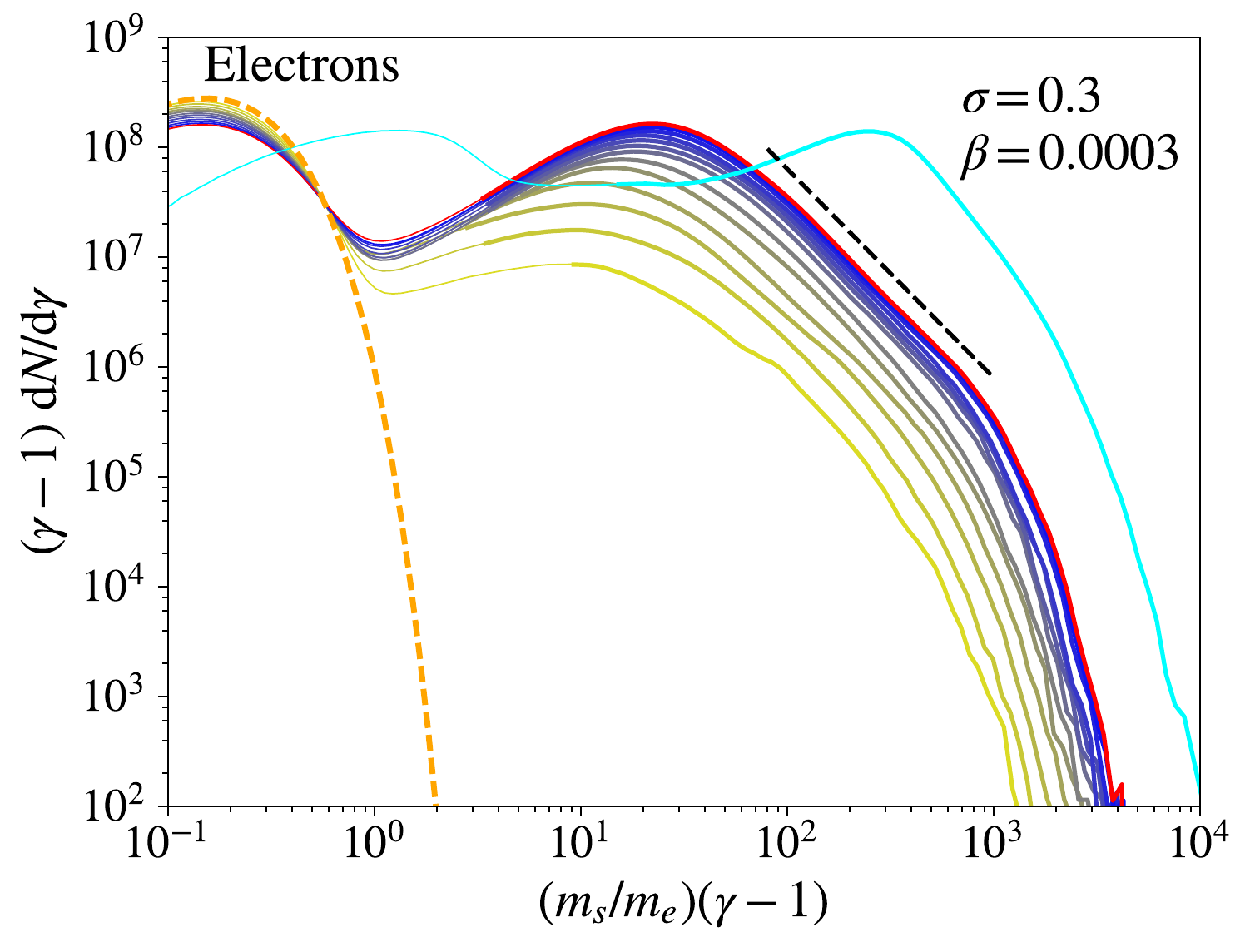}
}
\newline
{
	\includegraphics[width=\linewidth]{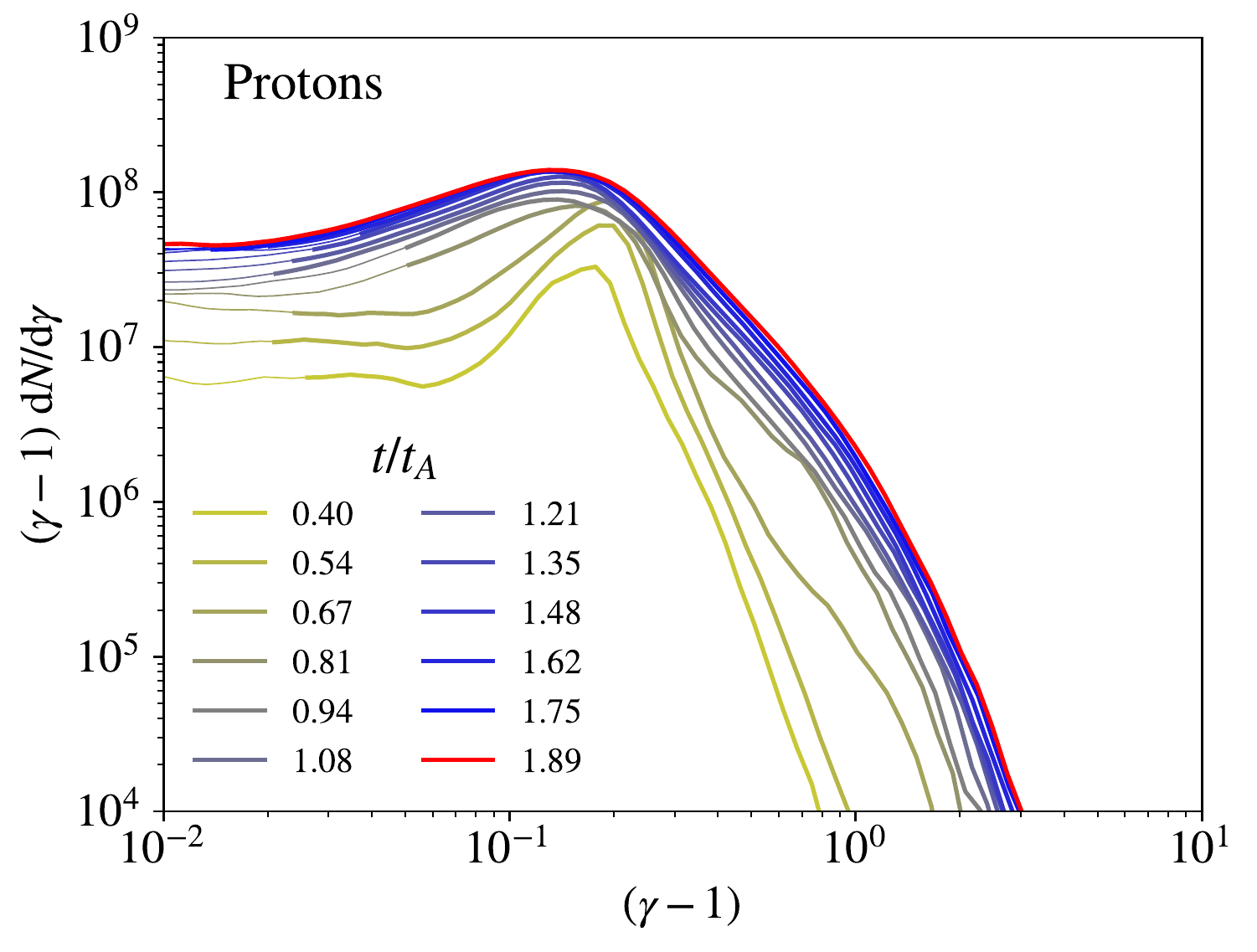}
}
\caption{Time evolution of the electron (top) and proton (bottom) energy spectra for the simulation shown in Figure \ref{lowbeta_threeplot}, with $\sigma=0.3$ and $\beta=3\times10^{-4}$ (run B1), corresponding to an initial proton thermal spread of $\theta_{i}=5 \times 10^{-5}$. The time sequence (from yellow to blue, with red marking the final time) is indicated in the bottom panel, in units of the Alfv\'enic crossing time $t_A=L_x/v_A$. In each panel, thicker lines indicate the energy range where the spectrum is mostly contributed  by particles in the reconnection region (by more than 75\%). In the top panel, the dashed orange line shows the initial electron Maxwellian for comparison, and the dashed black line represents a power law with slope $p=-d\log{N}/d\log{\gamma}=2.9$.
In the top panel, the proton spectrum at the final time (i.e., red curve in the bottom panel) is overplotted in cyan for comparison, with the horizontal axis rescaled by $m_i/m_e$. Since $m_s=m_e$ for electrons and $m_s=m_i$ for protons, the horizontal axis in the top panel represents the kinetic energy of each species, in units of the electron rest mass.}
\label{timespec}
\end{figure}

\subsection{Time Evolution of the Energy Spectra}
Having seen how the density evolves with time, and where and when various structures such as plasmoids and X-points form, we now examine the time evolution of the electron and proton energy spectra from the same simulation shown in Figure \ref{lowbeta_threeplot}.  We present the time evolution of the electron and proton spectra in the top and bottom panels of Figure \ref{timespec}, respectively. For both species, our spectra only include the particles that start in the ambient plasma (i.e., we exclude the contribution of the hot population that we set up in the current layer, whose properties depend on the initialization of the Harris sheet).  At each time, the spectrum includes all the particles in a fixed region delineated by  $\approx 0.2 L_{x}$ on each side of the current  sheet (bounded by the red lines in Figure \ref{regions}). On each curve, we also indicate with a thicker line the energy range where the fractional contribution of the reconnection region to the overall spectrum is greater than 75\%. We can thus identify the time evolution of the spectrum within the reconnection region itself, which we will refer to as the ``post-reconnection spectrum''.

In the top panel,  we present the evolution of the electron spectrum, which shows two components. The bump peaking at $\gamma-1 \approx 0.2$ is populated by the cold upstream electrons and is, in fact, well described by a Maxwellian distribution with the  temperature that we employ to initialize ambient electrons. The high-energy component, which peaks at $\gamma \approx 20$, is populated by particles that have been processed by reconnection. The high-energy component is consistent with a single non-thermal population having a power-law slope of $p=-d\log{N}/d\log{\gamma}=2.9$ that extends from the peak at $\gamma \approx 20$ up to $\gamma\approx1000$, where it cuts off exponentially. For reference, we show a power-law spectrum with an index of 2.9 with a dashed black line.  The power law starts right at the peak of the high-energy component, a common feature of magnetic reconnection (\citealt{sironi2014}; \citealt{melzani2014b, melzani2014}; \citealt{cerutti2012, cerutti2014, cerutti2017}; \citealt{guo2015}; \citealt{liguo2015};  \citealt{werner2018}). Moreover, the power-law index is established early on in the evolution of the electron energy distribution and does not change appreciably from $t=0.67\,t_{A}$ up to $t=1.89\,t_{A}$. The high-energy cutoff of the electron power law  steadily increases as larger plasmoids form and merge with each other or with the boundary island \citep{sironi2016}. 

In the bottom panel of Figure~\ref{timespec}, we show the time evolution of the proton energy spectrum.  The proton spectrum in the reconnection region resembles a power law at late times, similar to the electron spectrum. In the top panel, the proton spectrum at the final time ($t=1.89\,t_{A}$) is shown with a cyan line, with the horizontal axis scaled by $m_i/m_e$ in order to compare with the electron spectrum.  In other words, the horizontal axis in this figure indicates the kinetic energy of both species, in units of the electron rest mass energy. By comparing the thick cyan line for protons with the thick red line for electrons, we see that the proton mean energy in the reconnection region  is about an order of magnitude larger than the electron mean energy (see \citealt{rowan2017} and \citealt{werner2018} for a discussion of electron and proton heating in trans-relativistic reconnection). However, we also find that the proton spectrum has a steeper slope than the electron spectrum and that it spans a smaller range of energies.

The most dramatic difference between electron and proton spectra, though, is in their temporal evolution. At early times, the proton spectrum in the reconnection region is nearly monochromatic, with a pronounced peak at $\gamma-1 \approx  0.15 \approx  \sigma/2$, as expected from the characteristic kinetic energy of reconnection outflows (moving at $\sim v_A\sim c \,\sqrt{\sigma}$). Starting at $t\approx0.8\,t_A$, the spectrum develops a power-law-like tail. This transition occurs around the time when the two reconnection fronts interact across the periodic boundaries at a time in between panels (b) and (c) of Figure \ref{lowbeta_threeplot}, forming the large boundary island. This suggests that the interface between the reconnection outflows and the boundary island might be a promising source of non-thermal proton acceleration, as we further explore in Section \ref{mechanism}. We note that the development of a non-thermal proton distribution is not a peculiar consequence of our choice of triggering reconnection at the center of our domain. We observe the same evolution of the proton spectrum in untriggered runs, where the tearing mode is allowed to grow spontaneously.

In summary, protons develop a non-thermal tail only after $t\approx0.8\,t_A$, when the two reconnection fronts interact across the periodic boundaries. In contrast, electrons display a non-thermal component since early times. Although we only show here one particular choice of $\sigma$ and $\beta$, this trend holds across all of our low-$\beta$ simulations (the cases with $\beta$ approaching $\beta_{\rm max}$ are an exception, as we discuss below). These differences between the temporal evolution of electron and proton spectra point towards different acceleration mechanisms for the two species, as we discuss further in Section \ref{mechanism}. In particular, we will show that in low-$\beta$ cases, electrons are significantly accelerated at their first interaction with the layer by the non-ideal electric field at X-points. The early evidence for non-thermal electrons then comes from the fact that X-points appear since the earliest stages of evolution of the layer (see Figure \ref{lowbeta_threeplot}).

\begin{figure}[!h]
	\includegraphics[width =0.5\textwidth]{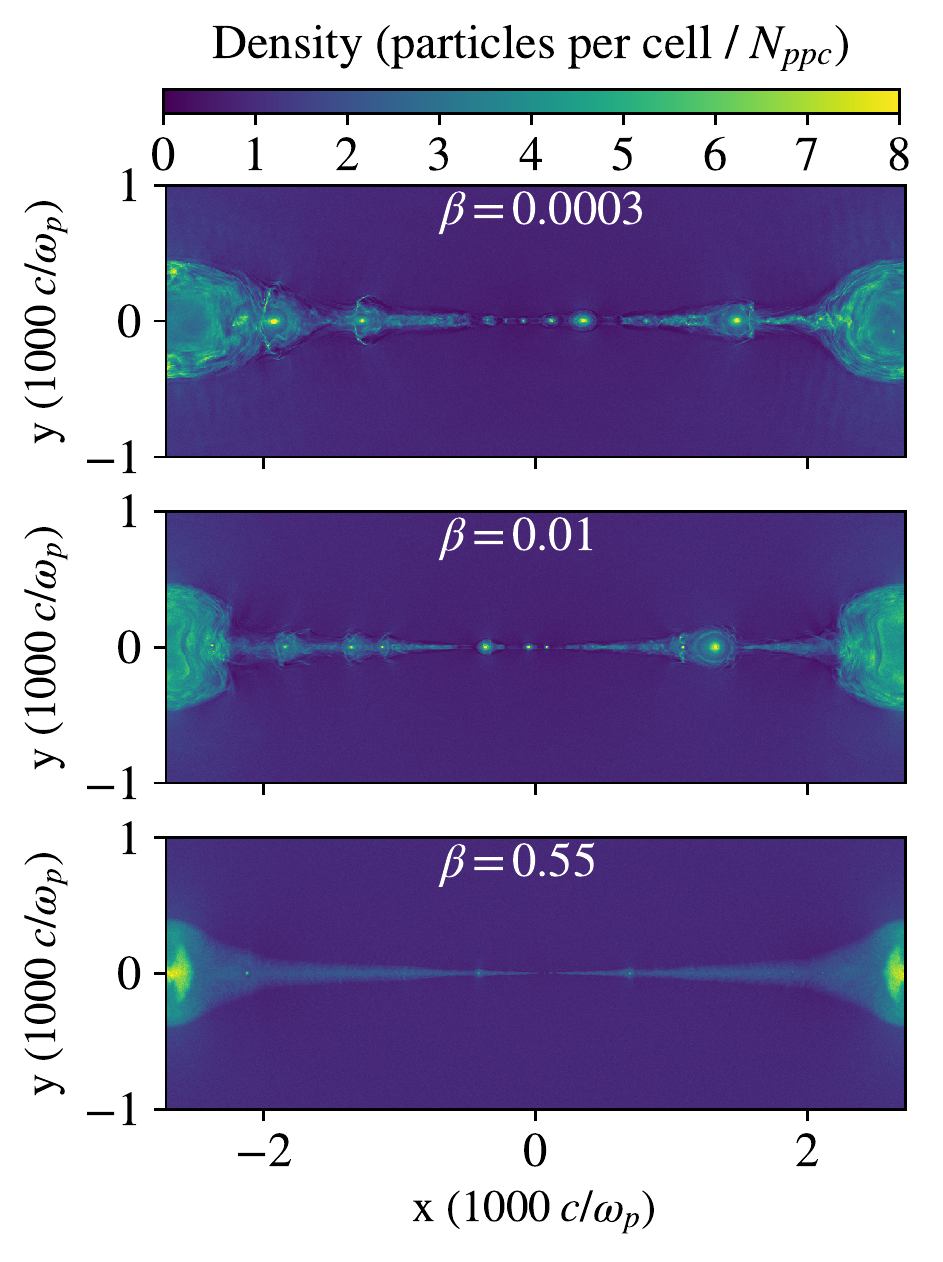}
	\caption{2D density structure  at $t\approx t_{A}$ for a suite of simulations with fixed $\sigma=0.3$ and varying $\beta$: $\beta=0.0003$ (top), $\beta=0.01$ (middle), and $\beta=0.55$ (bottom), corresponding to simulations B1, B5, and B8.  In the lowest-$\beta$ case (top), the reconnection layer is fragmented into numerous plasmoids separated by secondary X-points, whereas the highest-$\beta$ case (bottom) shows a smoother density profile along the reconnection outflows.}
    \label{sig_3_twoplot}
\end{figure}

\begin{figure}[!h]
	\includegraphics[width =0.5\textwidth]{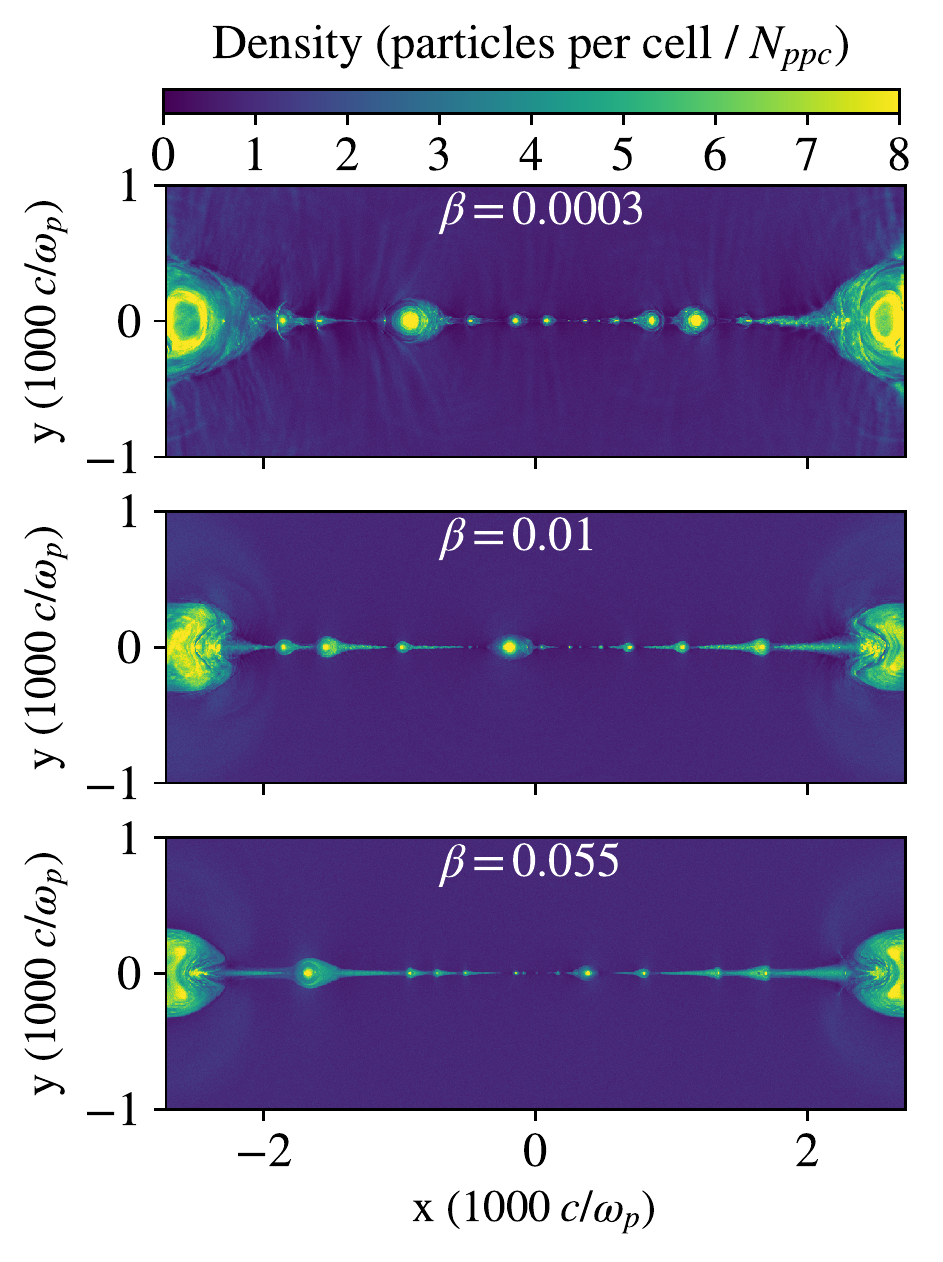}
	\caption{2D density structure  at $t\approx t_{A}$ for a suite of simulations with fixed $\sigma=3$ and varying $\beta$: $\beta=0.0003$ (top), $\beta=0.01$ (middle), and $\beta=0.055$ (bottom), corresponding to simulations D1, D4, and D6.  Secondary plasmoids form for all values of $\beta$, with larger plasmoids appearing in the lowest-$\beta$ simulation.}
    \label{sig3_twoplot}
\end{figure}

\section{The Role of $\beta$ in the Dynamics of the Reconnection Layer}\label{role_of_beta}
In this section, we illustrate how the dynamics in the reconnection layer depends on plasma beta and magnetization. First, we study the role of $\beta$ in the  development of 2D structures  in the reconnection region such as secondary plasmoids and X-points.  We then investigate the dependence on $\sigma$ and $\beta$ of the inflow rate (or equivalently, of the rate of magnetic field dissipation). In the next section, we will study the dependence on $\sigma$ and $\beta$ of the particle energy spectrum.

We show in Figure \ref{sig_3_twoplot} the 2D density structure of three simulations with fixed $\sigma=0.3$ and varying $\beta$: $\beta=0.0003$ (top), $\beta=0.01$ (middle), and $\beta=0.55$ (bottom).  In Figure \ref{sig3_twoplot} we show snapshots from three simulations with $\sigma=3$ (so, one order of magnitude higher than in Figure \ref{sig_3_twoplot}) and different $\beta$: $\beta=0.0003$ (top), $\beta=0.01$ (middle), and $\beta=0.055$ (bottom). For both figures, the snapshots are taken at $t\approx t_{A}$, after the reconnection fronts have reached the boundaries of the box.

In the low-$\sigma$ case with $\sigma=0.3$, we see a clear difference in the structure of the current sheet between low- and high-$\beta$ simulations. At low $\beta$, the current layer is pinched by the secondary tearing mode at multiple locations along the sheet, resulting in numerous secondary X-points and plasmoids. In contrast, the highest-$\beta$ case, which is close to $\beta_{\rm max}\approx 1/4\sigma\simeq 0.8$, displays a smooth density profile in the reconnection outflows, with only marginal evidence for two secondary plasmoids. No prominent X-points are detected at high $\beta$, with the exception of the primary X-point located at the center of the layer, resulting from our initial perturbation of the current sheet.

In the high-$\sigma$ simulations with $\sigma=3$, the dependence on $\beta$ is less pronounced.  We do, however, see that the lowest-$\beta$ case has larger plasmoids and that its current layer is broken up into distinct high-density plasmoids, separated by low-density regions.  In comparison, the highest-$\beta$ simulation in the bottom panel (with $\beta$ approaching $\beta_{\rm max}\approx 1/4\sigma\simeq 0.08$) still presents several secondary plasmoids, but the density profile in between neighboring plasmoids is smoother than at lower $\beta$. In other words, the density contrast between secondary plasmoids and X-points seems to get reduced with increasing $\beta$.

In summary, the fragmentation of the current sheet into secondary plasmoids separated by secondary X-points becomes increasingly pronounced at lower $\beta$ (for fixed $\sigma$) and at higher $\sigma$ (for fixed $\beta$; see also \citealt{sironi2016}, for the same conclusion in the ultra-relativistic regime $\sigma\gg1$).  It is likely that these structural differences in the appearance of the reconnection layer play a key role in whether efficient particle acceleration occurs, as we will discuss in Section \ref{mechanism}.

\subsection{Reconnection Rate}
In the whole range of $\sigma$ and $\beta$ we investigated in this work, we calculate the mean inflow rate, which corresponds to the rate of magnetic field dissipation (i.e., to the so-called ``reconnection rate''). At each time, we compute the spatial average of the $y-$component of the flow velocity in a region close to the center of the domain, covering the range  $|y| \lesssim 400\; c/\omega_{p}$ and $|x| \lesssim 1000\; c/\omega_{p}$. This area is sufficiently large that it allows to obtain a proper estimate of the steady-state inflow rate and it is chosen to exclude the boundary island, which artificially inhibits the plasma inflow rate in its vicinity. 

\begin{figure}[!h]
{
\includegraphics[width=\linewidth]{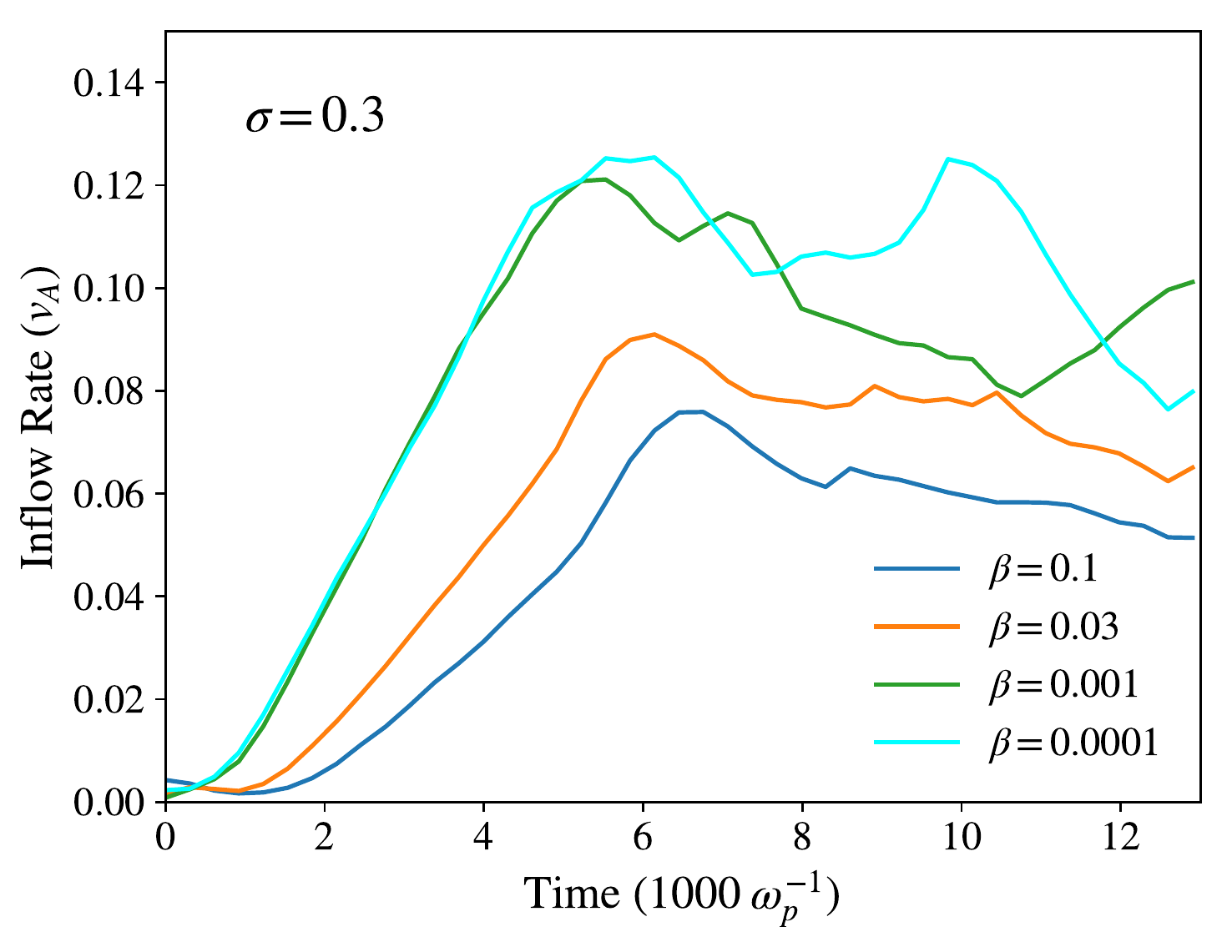}
\caption{\label{sigpoint3_inflow_rates}Time evolution of the inflow rate, in units of the upstream Alfv\'en speed, for four different simulations at fixed $\sigma=0.3$ and varying $\beta$ (simulations B0, B2, B6, and B7, in order of increasing $\beta$).  The inflow speed tends to decrease at higher $\beta$.}
}
\end{figure}

In Figure \ref{sigpoint3_inflow_rates}, we show the temporal evolution of the inflow rate for four representative simulations that have a fixed $\sigma=0.3$ and range in $\beta$ over three orders of magnitude. The inflow speed is measured in units of the upstream Alfv\'{e}n velocity $v_A$. At early times ($\omega_{p}t\lesssim 5000$), the inflow rate steadily increases, as the reconnection fronts move away from the center of the domain, and the region of inflowing plasma  extends further and further in both the $x$ and $y$ directions. 
After the reconnection rate reaches its peak, it settles around a constant value (with only a slight decrease at later times). Eventually, the boundary island could grow large enough to inhibit the inflow of particles and magnetic flux and the reconnection rate would artificially drop to zero. Figure \ref{sigpoint3_inflow_rates} suggests that, for the timespan covered by our simulations, the computational domain is sufficiently large to properly capture the steady state of reconnection, without artificial effects from the periodic boundaries.

At low $\beta$, the inflow rate displays significant fluctuations. After the peak, the reconnection rate drops. This is due to the fact that 
the first secondary plasmoids tend to form around the center of the box and their pressure slightly inhibits the inflow of surrounding upstream plasma. Once the plasmoids get advected by the field tension towards the boundary island, the upstream plasma can freely flow into the layer, which explains the second peak in the reconnection rate (at $\omega_{p}t\sim 9500$ for $\beta=10^{-4}$ and at $\omega_{p}t\sim 13000$ for $\beta=10^{-3}$). These oscillations in the temporal profile of the inflow rate are observed for all our low-$\beta$ cases.

\begin{figure}[t]
\includegraphics[width =0.5\textwidth]{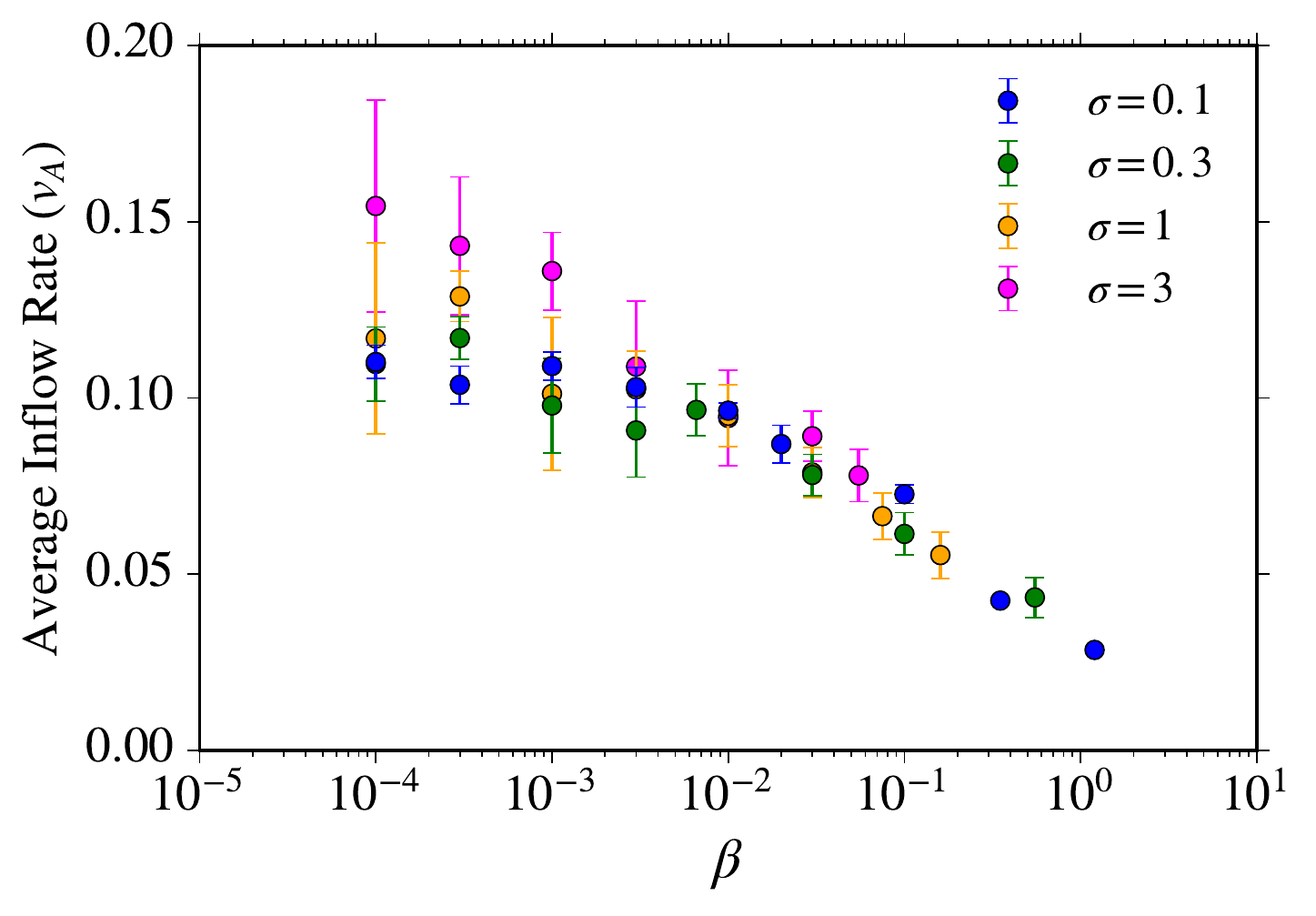}
	\caption{Temporal averages of the inflow rate as a function of $\sigma$ and $\beta$, in units of the upstream Alfv\'en velocity.  The error bars indicate the standard deviation, which is larger at low $\beta$ for the copious formation of secondary plasmoids.}
    \label{inflow_rates_over_Va}
\end{figure}

From  Figure \ref{sigpoint3_inflow_rates}, it is clear that the inflow rate is nearly independent of $\beta$ in the low-$\beta$ regime, but it tends to decrease at higher $\beta$. This is further confirmed by Figure~\ref{inflow_rates_over_Va}, where we present, as a function of $\sigma$ and $\beta$, the mean reconnection rate, averaged from the peak time to a time $3000 \; \omega_{p}^{-1}$ ($\sim 0.3 \,t_{A}$) after the peak time, when the reconnection process is  steadily active. The error bars in Figure \ref{inflow_rates_over_Va} indicate the standard deviation, which is larger at lower $\beta$, where the copious formation of secondary plasmoids causes pronounced oscillations in the inflow rates, as we have discussed above.

From Figure \ref{inflow_rates_over_Va}, we see that the inflow rate for $\beta\gtrsim 10^{-2}$ is nearly independent of $\sigma$, but it gets lower and lower for increasing $\beta$. This behavior was noted in MHD simulations by \citealt{ni2012}, and in PIC simulations by \citealt{rowan2017}. For $\beta\lesssim 10^{-2}$, the inflow rate is nearly $\beta$-independent (with the exception of $\sigma=3$), and tends to increase with $\sigma$ when approaching the relativistic regime $\sigma\gtrsim 1$ (see  \citealt{sironi2016} for the dependence of the inflow rate on magnetization in the ultra-relativistic regime $\sigma\gg1$). The low-$\beta$ limit at $\sigma\lesssim1$ is consistent with a fixed value of the reconnection rate, of order $\sim 0.1\,v_{A}$. 
As we further discuss in the next two Sections, the dependence of the inflow velocity on $\beta$ and $\sigma$ will be mirrored by the magnitude of the electric field in the reconnection region, which in turn impacts the rate of particle acceleration.

\section{Dependence on  $\beta$ and $\sigma$ of the Electron Energy Spectra}\label{energy_spectra}
In this section, we investigate the role of $\sigma$ and $\beta$ on the physics of non-thermal particle acceleration, with focus on electron acceleration. We first describe how we characterize the non-thermal electron energy spectrum in the reconnection region, finding that it can be generally modeled as a power law. We quantify how the slope of the power law and the electron acceleration efficiency depend on $\beta$ and $\sigma$. We also discuss an additional high-energy component that appears for $\beta$ approaching $\beta_{\rm max}$ in both electron and proton spectra.

\subsection{Characterizing the Electron Energy Spectra}\label{chara}
\begin{figure}[!t]
	\includegraphics[width =0.5\textwidth]{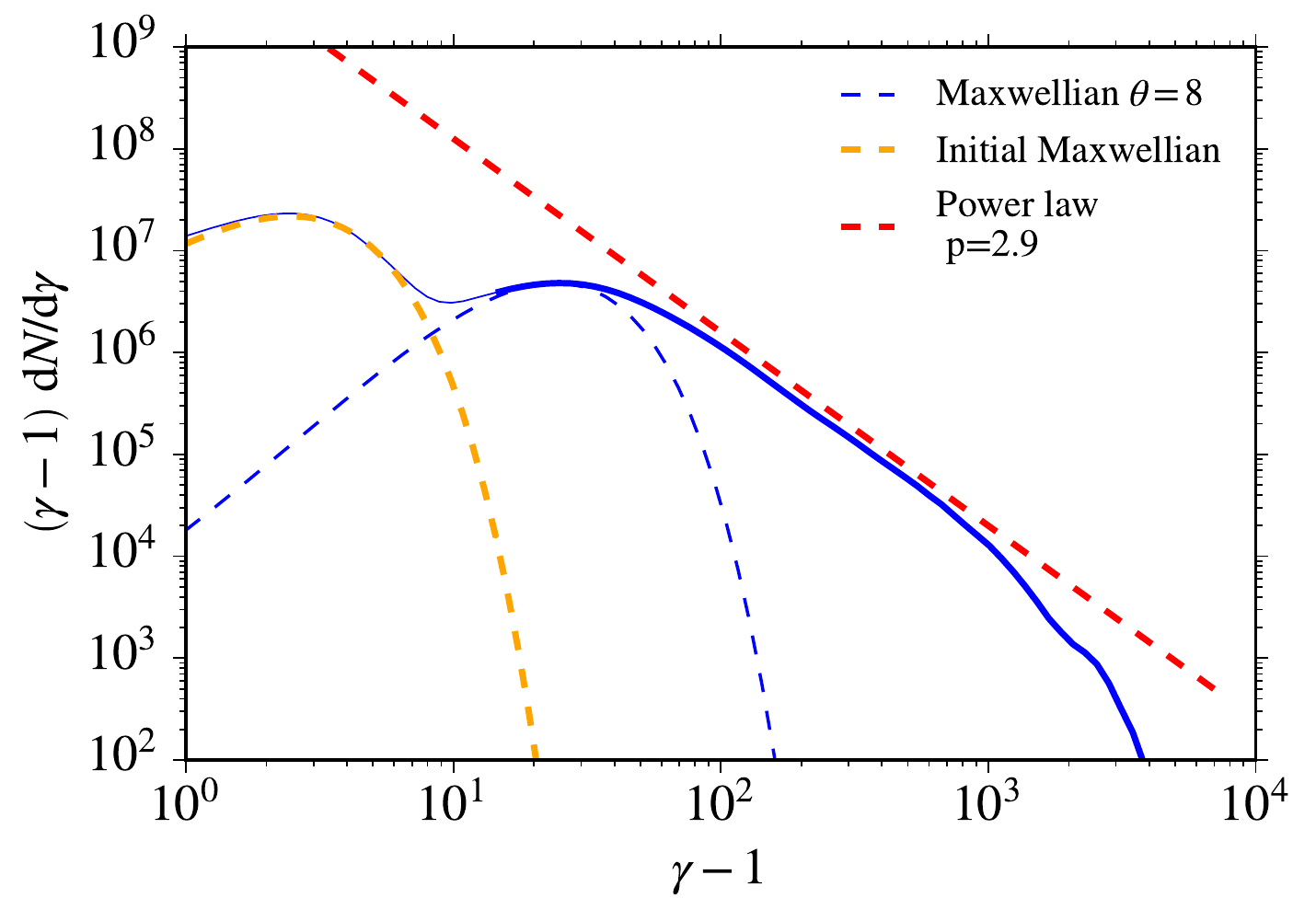}
	\caption{Electron spectrum for a simulation with $\sigma=0.3$ and $\beta=0.003$ (simulation B3) taken at $t=2\,t_{A}$.  The solid blue line shows the overall spectrum in the slab delimited by the red lines in Figure~\ref{regions}, and the thick blue line marks the energy range where the spectrum is mostly contributed by the reconnection region (yellow area in Figure~\ref{regions}).  The dashed blue line shows the Maxwellian fit to the peak of the spectrum in the reconnection region, the red dashed line shows the best-fitting power law, and the orange dashed line depicts the  Maxwellian distribution initialized in the inflow region.}
\label{spec_fit_ex}
\end{figure}

As an illustrative example of how we characterize the properties of electron spectra, we show in Figure~\ref{spec_fit_ex} the electron energy distribution for a simulation with $\sigma=0.3$ and $\beta=0.003$. The solid blue line depicts the electron spectrum measured in a slab with $|y|\lesssim 1000\,c/\omega_{p}$, as delimited by the red lines in  Figure \ref{regions}. As before, the portion of the blue curve plotted with a thicker blue line in Figure~\ref{spec_fit_ex} indicates the energy range where the reconnection region contributes more than 75\%. The dashed orange line shows the Maxwellian distribution initialized in the inflow region, demonstrating that the low-energy bump in the electron spectrum is populated by particles that have yet to experience the reconnection process. The high-energy component is a genuine by-product of the reconnection physics. It can be modeled as a power law (compare with the dashed red line, that has a slope of $p=2.9$).

In addition to the power-law slope, we quantify the efficiency of reconnection in producing non-thermal particles, by employing the following strategy. We isolate the spectrum of the reconnection region and fit its peak with a relativistic Maxwellian $f_{\rm MB} (\gamma,\theta)$ (shown in the dashed blue line in Figure~\ref{spec_fit_ex}), which has a dimensionless temperature $\theta=kT/m_e c^2\approx8$).  The spectrum exceeds the Maxwellian distribution for $\gamma > \gamma_{\rm{pk}}$ where $\gamma_{\rm{pk}}$ denotes the peak of the spectrum.  Based on this, we can quantify
the efficiency of electron acceleration by integrating the excess of the electron spectrum with respect to the best-fitting Maxwellian for 
$\gamma > \gamma_{\rm{pk}}$, normalized to the overall energy content of the spectrum. Thus we define the non-thermal acceleration efficiency $\epsilon$ as
\begin{equation}
\epsilon= \frac{\int_{\gamma_{\rm{pk}}}^{\infty}(\gamma-1)[ \frac{dN}{d\gamma} -f_{\rm MB}(\gamma,\theta)]d\gamma}{\int_{\gamma_{\rm pk}}^{\infty}(\gamma-1) \frac{dN}{d\gamma}d\gamma}~,
\label{efficiency_eqn}
\end{equation}
where $\theta$ is the best-fit dimensionless temperature. In Section \ref{sec:5.4}, we will employ this strategy to characterize how the non-thermal acceleration efficiency and the electron power-law slope depend on plasma beta and magnetization by taking the electron spectrum at $t\approx 2\,t_{A}$, when the spectral shape has saturated.

We conclude this subsection with two cautionary remarks. First, as we discuss in Section \ref{sec:5.2}, the electron spectrum softens with increasing $\beta$.  This makes the determination of the electron power-law slope and non-thermal efficiency less accurate for higher values of $\beta$. Second, as we describe in Section \ref{betamax}, a peculiarity of the extreme cases with $\beta\sim \beta_{\rm max}$ is the presence of a separate high-energy spectral component, containing a few percent of particles. As we discuss below, the particles belonging to this additional component experience a different energization process than the bulk of electrons accelerated by reconnection. For this reason, we neglect this additional component when characterizing the non-thermal acceleration efficiency. In practice, for the small set of simulations with $\beta\sim \beta_{\rm max}$, we identify the Lorentz factor where the additional component starts, and we take this as an upper limit in Equation \ref{efficiency_eqn}, rather than integrating up to infinity.

\begin{figure}[!b]
	\includegraphics[width =0.5\textwidth]{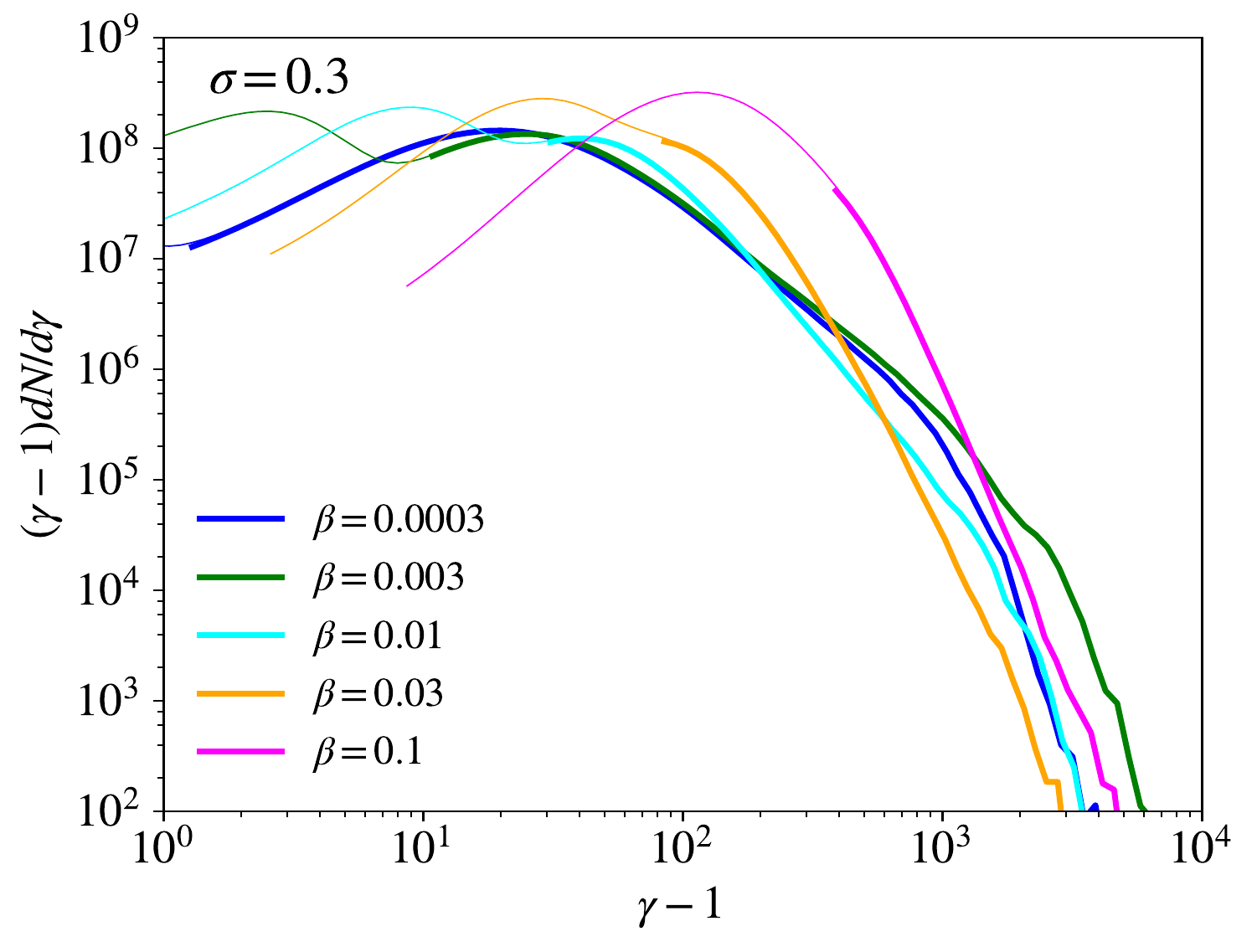}
	\caption{Electron spectra for fixed $\sigma=0.3$ and varying $\beta$, as indicated in the legend (simulations B1, B3, B5, B6, and B7), calculated at $t\approx 2\,t_{A}$.  At low $\beta$, the spectral shape converges (e.g., the blue and green curves have the same spectral slope), but as $\beta$ increases, the power law steepens significantly. Thicker lines indicate post-reconnection spectra.}
    \label{sig_3_betas_lecs}
\end{figure}

\subsection{Dependence on $\beta$  and $\sigma$ of Electron Energy Spectra}\label{sec:5.2}
In this section, we present a few representative electron energy spectra to illustrate their dependence on $\beta$  and $\sigma$. All  the spectra are measured at $t=2\, t_{A}$. As usual, thicker lines indicate the spectral range dominated by particles residing in the reconnection region. 

In Figure~\ref{sig_3_betas_lecs}, we show five electron spectra from simulations with $\sigma=0.3$ and a wide range of $\beta$. 
At low beta ($\beta\lesssim 3\times 10^{-3}$), the post-reconnection spectra, shown in blue and green thick lines, peak at $\gamma\sim 20$, regardless of $\beta$. This is consistent with the results of \citet{rowan2017}, who showed that, at sufficiently low $\beta$, the reconnection process converts a fixed amount of magnetic energy into electron energy, regardless of the initial $\beta$. In addition, Figure~\ref{sig_3_betas_lecs} shows that  the shape of the post-reconnection spectrum is nearly the same for all values of $\beta\lesssim 3\times10^{-3}$. Both the power-law slope and the high-energy cutoff are insensitive to the plasma-$\beta$, in the range $\beta\lesssim 3\times 10^{-3}$. The small degree of variation in the slope and high-energy cutoff between the cases with $\beta=3\times10^{-3}$ and $\beta=3\times10^{-4}$ is due to the stochastic nature of the plasmoid chain. In fact, in the $\beta=3\times10^{-3}$ simulation, a sequence of consecutive mergers leads to the formation of an unusually large secondary plasmoid. Each merger is accompanied by efficient electron acceleration and the peculiar merger history of the $\beta=3\times10^{-3}$ case results in the high-energy slope being slightly harder and extending to higher energies than in other simulations with comparable $\beta$.

At higher beta values, the separation between the thermal peak of inflowing particles and the post-reconnection component shrinks, since the energy content in magnetic fields available for dissipation becomes an increasingly smaller fraction of the plasma thermal energy. In these high-$\beta$ cases, the spectrum of the plasma that has undergone reconnection can only be identified using our mixing criterion, which is based on the spatial distinction between the upstream flow and the post-reconnection region, rather than a spectral distinction. When $\beta$ increases beyond $sim 10^{-2}$, we find that the power-law slope steadily steepens and the overall spectrum  eventually resembles a single Maxwellian distribution. This trend holds for all the magnetizations we have investigated, as we further discuss in Section \ref{sec:5.4}.

\begin{figure}[!h]
	\includegraphics[width =0.5\textwidth]{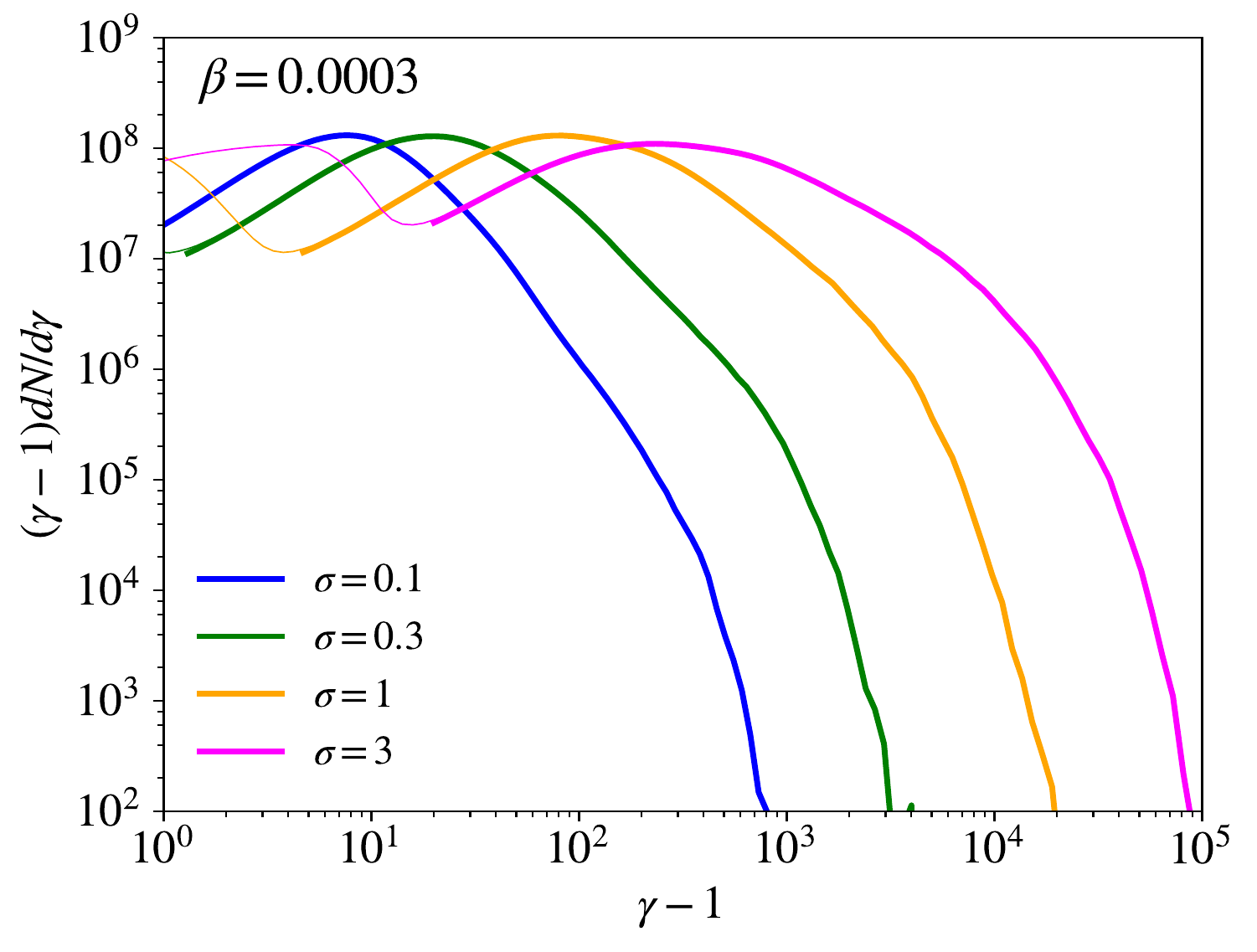}
	\caption{Electron spectra for a set of simulations with fixed $\beta=3\times 10^{-4}$ and varying $\sigma$, as indicated in the legend (simulations A1, B1, C1, and D1), measured at $t\approx 2\,t_{A}$.  As $\sigma$ increases, the spectra broaden and the slope hardens.}
    \label{beta0003_spec}
\end{figure}

\begin{figure}[!h]
	\includegraphics[width =0.5\textwidth]{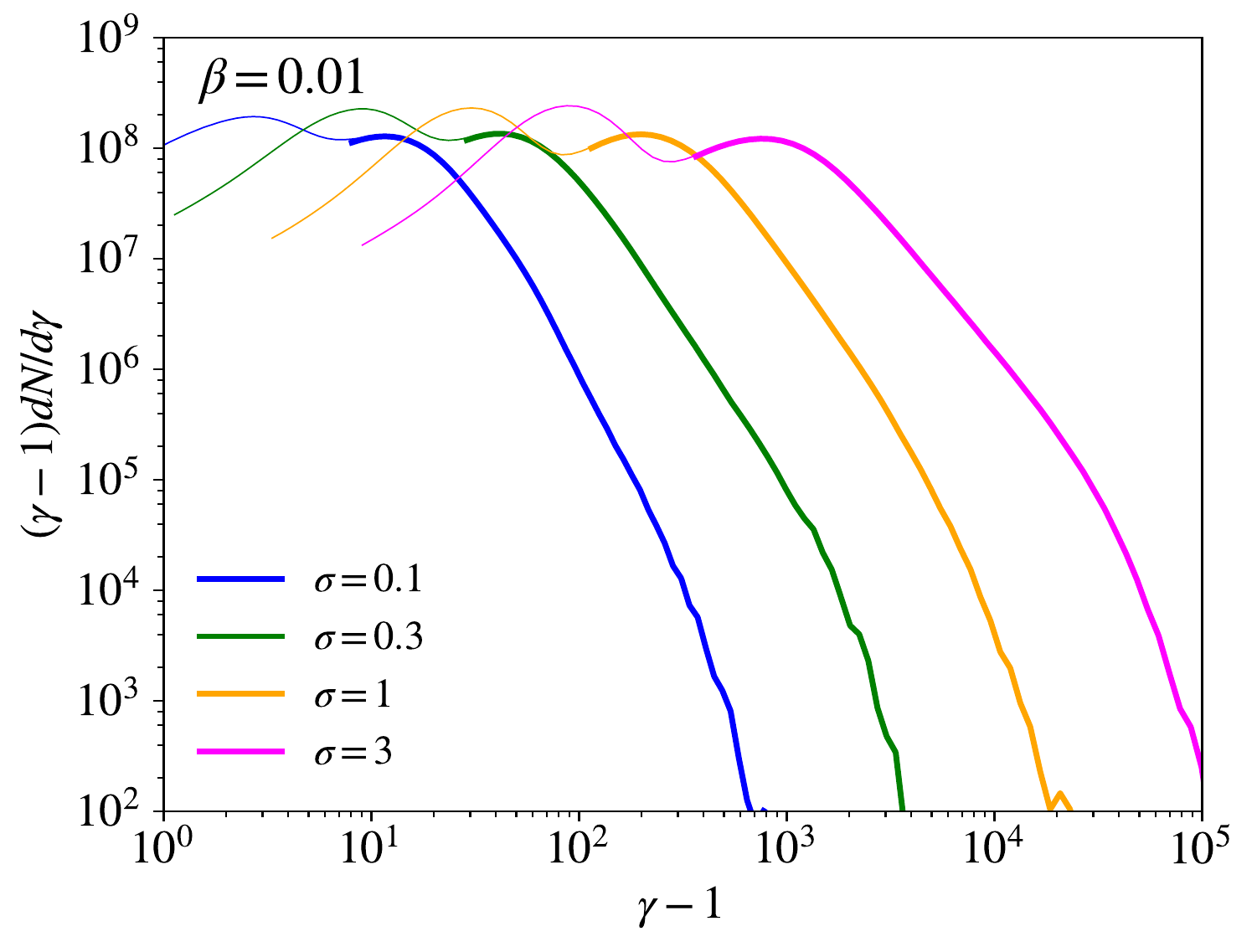}
	\caption{Electron spectra for a set of simulations with fixed $\beta=0.01$ and varying $\sigma$, as indicated in the legend (simulations A4, B5, C4, and D4), measured at $t\approx 2\,t_{A}$. This choice of $\beta$ is the same as in the work by  \citet{werner2018}.}
    \label{beta01_spec}
\end{figure}

In Figures~\ref{beta0003_spec} and \ref{beta01_spec}, we explore how the electron spectra change when varying $\sigma$, at fixed $\beta$. As $\sigma$ increases, the amount of magnetic energy available for dissipation increases, which accounts for the shift to higher energies in the peaks of post-reconnection spectra. 
More interestingly, for $\beta=3\times10^{-4}$ (Figure~\ref{beta0003_spec}), we see that the post-reconnection spectrum becomes significantly harder with increasing $\sigma$.  The same is observed for $\beta=0.01$ (Figure~\ref{beta01_spec}), although the trend is not as prominent.

This trend --- of harder spectral slopes for higher $\sigma$ --- has been already discussed by \citet{werner2018}. In fact, the four simulations  in Figure~\ref{beta01_spec} have the same physical parameters as in \citet{werner2018}, where the dependence on $\sigma$ was investigated for the specific case of $\beta=0.01$.
In \citet{werner2018}, the electron power-law slopes for $\sigma=0.1, \; 0.3, \; 1,$ and $ 3$ were measured to be 4.0, 3.3, 2.8, and 2.4, respectively.  For these same values of $\sigma$ and $\beta$, we measure power-law indices of 4.3, 3.8, 3.6, and 3.2, i.e., we find that our spectra are systematically softer than in \citet{werner2018}. We attribute this discrepancy to the combination of two effects. First, our simulation domain for $\beta=0.01$ is about five times larger than that of \citet{werner2018}. As we discuss in Appendix \ref{boxsize}, larger domains systematically lead to steeper electron spectra. Second, as we describe in Appendix \ref{untriggered}, we find appreciable differences in the hardness of the electron spectrum between our setup, where reconnection is triggered in response to a large-scale perturbation, and the untriggered case, where the reconnection spontaneously evolves from particle noise. In particular, the untriggered setup generally leads to harder electron spectra. We have verified that we  recover the power-law slopes quoted by  \citet{werner2018} in the case of untriggered simulations with the same box size that they employ.

\subsection{The Additional High-Energy Component at $\beta \sim \beta_{\rm max}$}\label{betamax}
\begin{figure}[!h]
\includegraphics[width =0.5\textwidth]{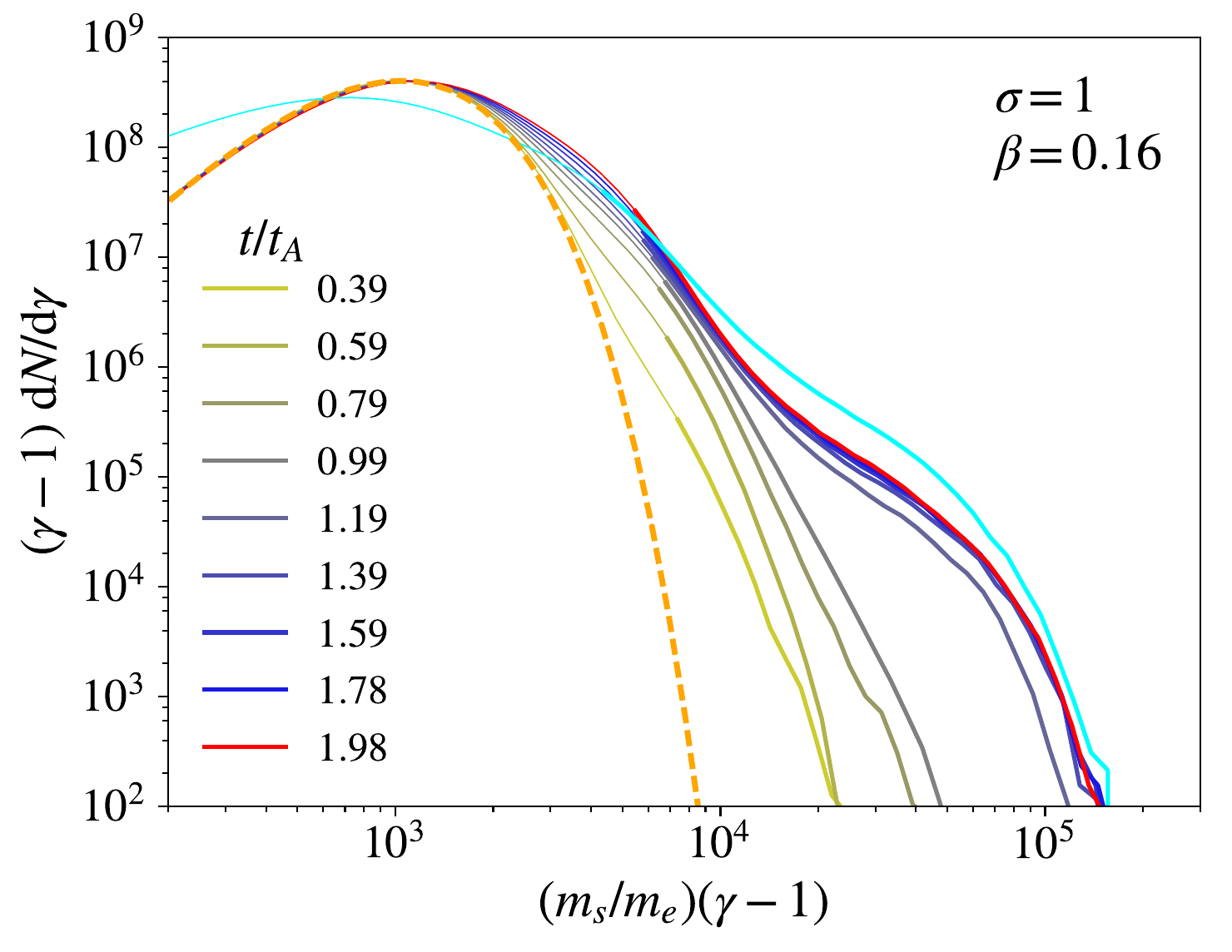}
\caption{Time evolution of the electron spectrum in the simulation with $\sigma=1$ and $\beta=0.16$ (simulation C7) that shows the strongest evidence for the additional high-energy component seen as $\beta\rightarrow\beta_{\rm max}$. We show the upstream electron Maxwellian with a dashed orange line. The proton spectrum at the final time is shown with the cyan line, with the horizontal axis rescaled by the mass ratio for comparison. Time is in units of the Alfv\'enic crossing time $t_A=L_x/v_A$.}
\label{sig1_timespec}
\end{figure}

A peculiarity of the extreme cases with $\beta\sim \beta_{\rm max}$, which are marked with an asterisk in Table 1, is the presence of a separate high-energy spectral component emerging at late times. In Figure \ref{sig1_timespec}, we show  the temporal evolution of the electron spectrum in the simulation  that shows the strongest evidence for this additional component (i.e., the case with $\sigma=1$ and $\beta=0.16$).  

At early times ($t\lesssim t_{A}$), the high-energy part of the spectrum is very steep and is barely distinguishable from the upstream Maxwellian. At later times ($t\gtrsim t_{A}$), an additional component appears at high energies. It develops around the time when the boundary island is formed by the interaction of the two reconnection fronts across the periodic boundaries. As we show in Section \ref{mechanism}, the electrons belonging to this additional high-energy component are accelerated by bouncing between the reconnection outflow and the boundary island, in a process reminiscent of the Fermi mechanism. This additional high-energy component is a generic outcome of high-$\beta$ reconnection. In particular, it is not an artificial by-product of our choice of a triggered reconnection setup, since it also appears in untriggered simulations, as we show in Appendix \ref{untriggered}. 

In Figure \ref{sig1_timespec}, we also show with a cyan line the proton spectrum at the final time. We find that the proton spectrum displays a similar high-energy component, with just a slightly higher normalization. In other words,  electrons and protons are subject to the same  acceleration mechanism.
In retrospect, this is not surprising: in the limit that $\beta$ approaches $\beta_{\rm max}$, the upstream protons become trans-relativistic ($\theta_{i}=0.2$ for the case we show). Because the upstream electrons are also relativistic, the two species have comparable Larmor radii, and are then expected to be accelerated in a similar fashion.

\subsection{Dependence of the Power-Law Slope and Acceleration Efficiency on $\beta$  and $\sigma$ }\label{sec:5.4}
\begin{figure}[!h]
\includegraphics[width =0.5\textwidth]{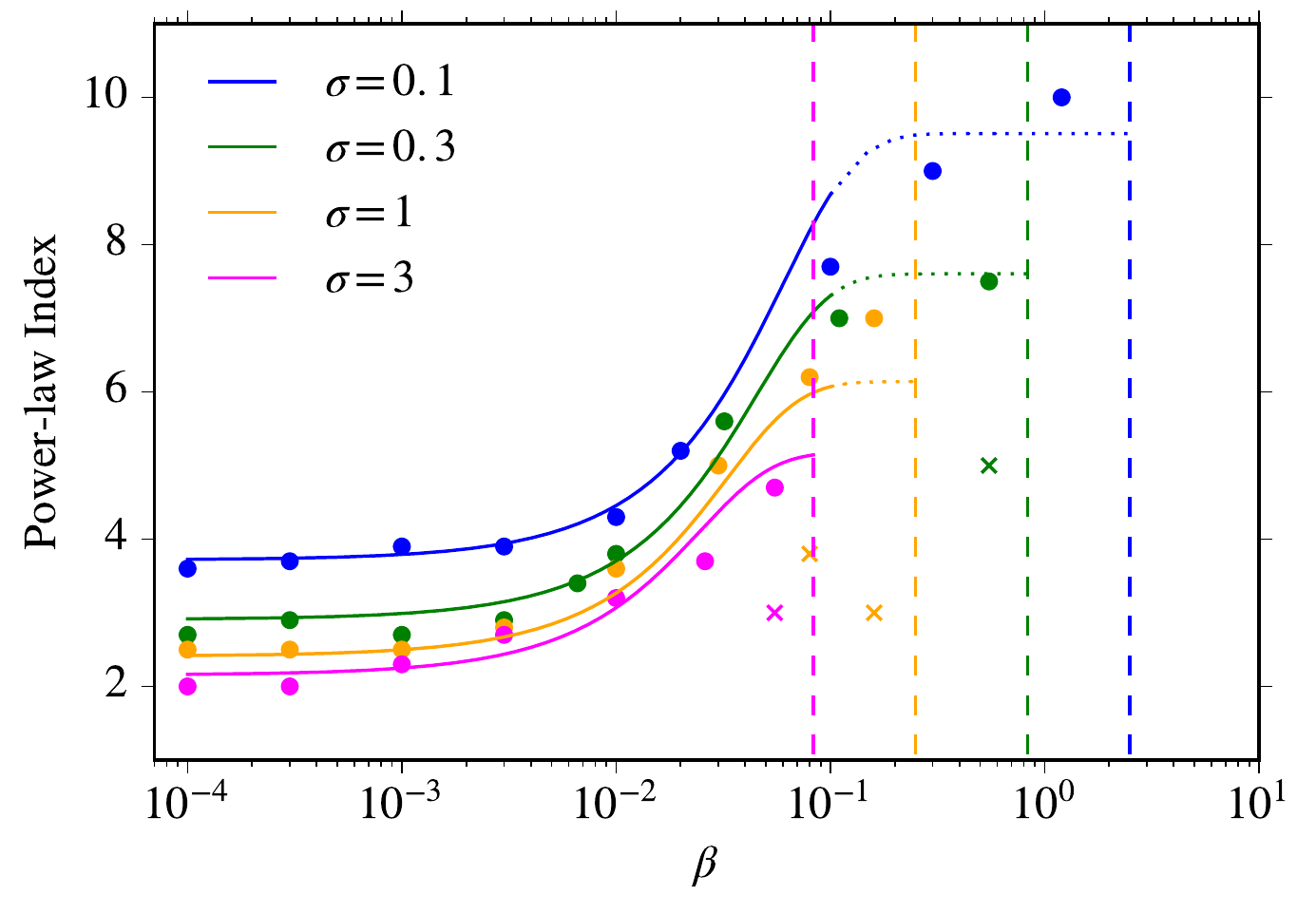}
\caption{Electron power-law slope as a function of $\beta$ (horizontal axis) and for different values of $\sigma$ (different colors, as indicated in the legend).  The power-law indices of the main non-thermal component (i.e., the one starting from the thermal peak) are depicted with filled circles, while the power-law indices of the additional high-energy bump that appears for $\beta\sim \beta_{\rm max}$ (i.e., in the simulations marked with an asterisk in Table 1) are indicated with crosses.  The solid lines show our empirical fit in Equation \ref{initial_tanfit}. Beyond $\beta\sim 0.1$, the electron spectra become very steep and so our estimates are less robust (for this reason, our fitting curves for $\beta\gtrsim 0.1$ are plotted as dotted lines).  The values of $\beta_{\rm{max}}$ for each $\sigma$ are indicated with vertical dashed lines.}
\label{powerlaw_fit}
\end{figure}

\begin{figure}[!h]
\includegraphics[width =0.5\textwidth]{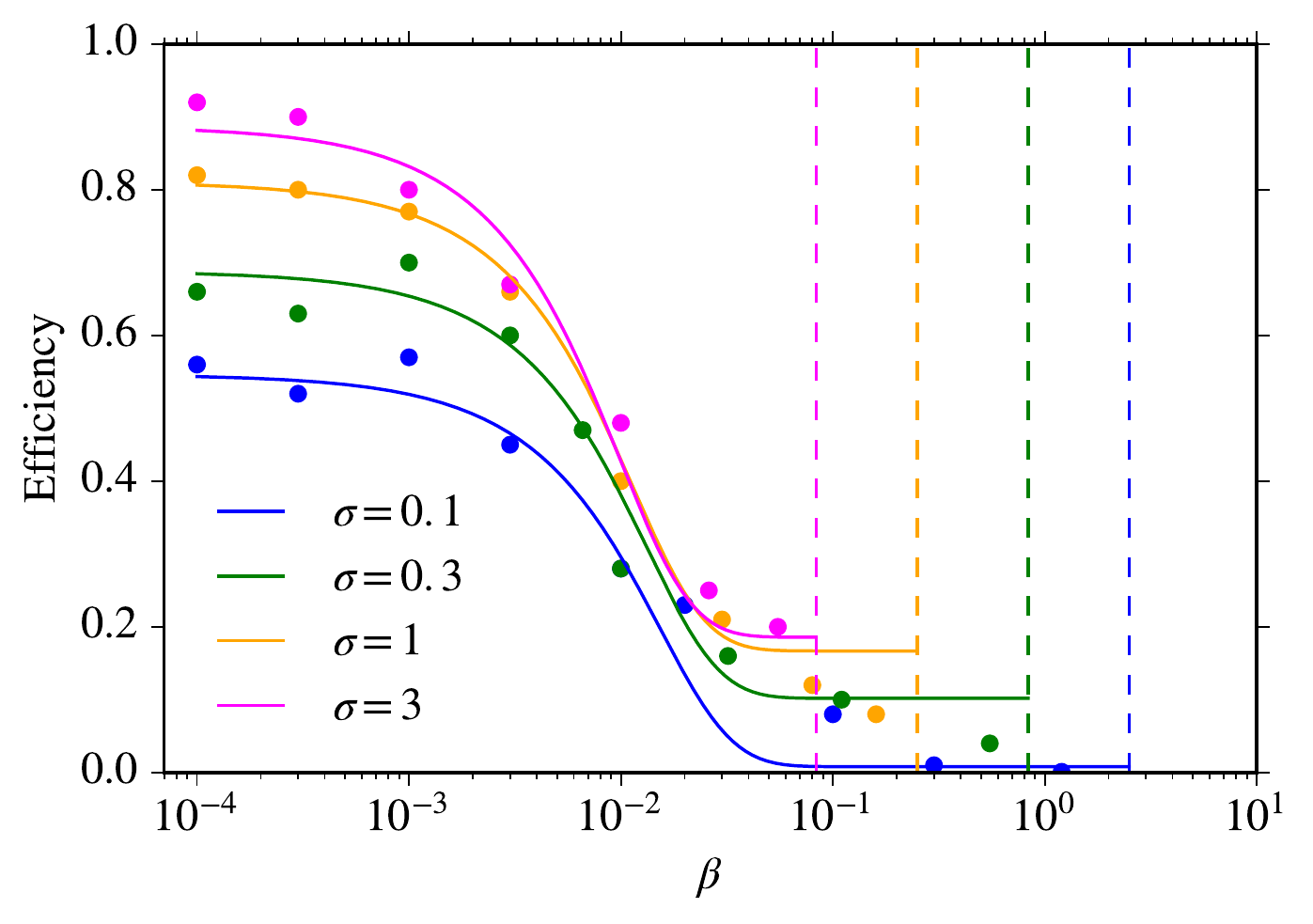}
\caption{Electron non-thermal acceleration efficiency $\epsilon$ as a function of $\beta$ (horizontal axis) and for different values of $\sigma$ (different colors, as indicated in the legend). The solid lines show our empirical fit in Equation \ref{fiteff}. For each $\sigma$, the solid lines extend up to maximum allowed $\beta$, i.e., $\beta_{\rm max}=1/4\sigma$.}
\label{efficiency_fit}
\end{figure}

In this Section, we summarize our results on the dependence of the electron energy spectrum on magnetization and plasma beta. In Figure \ref{powerlaw_fit}, we show how the electron power-law slope depends on $\beta$ and $\sigma$, and in Figure  \ref{efficiency_fit}, we present the dependence on $\beta$ and $\sigma$ of the efficiency of non-thermal electron acceleration, as defined in Equation \ref{efficiency_eqn}. 

In Figure \ref{powerlaw_fit}, filled circles indicate the slope of the main component of accelerated electrons, while crosses represent the slope of the additional component that emerges for $\beta\approx \beta_{\rm max}$ at late times.  We also show the values of $\beta_{\rm max}$ for each $\sigma$ with vertical dashed lines. When focusing on the filled circles, two trends are evident. First, at fixed $\beta$, the power-law slope is harder for higher $\sigma$ (see also \citealt{werner2018}). Second, at fixed $\sigma$, the slope is independent of $\beta$ for $\beta \lesssim 3 \times 10^{-3}$, but it increases at higher values of $\beta$, eventually resulting in a non-thermal tail that is too steep to be distinguishable from a Maxwellian distribution.

It is possible to express the results of these PIC calculations in an analytical form and to employ these prescriptions as a sub-grid model in larger scale simulations of trans-relativistic plasmas. To this end, we empirically fit the combined dependence of the electron slope $p$ on plasma $\beta$ and magnetization $\sigma$ using the functional form

\begin{equation}
p = A_{p} + B_{p} \tanh{(C_{p}\beta)}~~,
\label{initial_tanfit}
\end{equation}
where
\begin{equation}
\begin{aligned}
A_{p}=1.8 + 0.7/\sqrt{\sigma}~,~B_{p}=3.7\,\sigma^{-0.19}~,~C_{p}=23.4\,\sigma^{0.26}   ~.   
\end{aligned}
\label{powerlaw_fit_params}
\end{equation}
We show this fit with solid lines in Figure \ref{powerlaw_fit}. For $A_p$, we have employed an expression similar to \citet{werner2018}, which properly captures the $\sigma$-dependence of our results in the limit $\beta\ll1$ \footnote{In principle, for $\beta\lesssim 3\times 10^{-3}$ the slope can take on an even simpler form and be written as $p\simeq 1.8 + 0.7/\sqrt{\sigma}$.}. Specifically, in this low-$\beta$ regime, the expression for the electron power-law slope approaches $p\simeq 1.8$ for $\sigma\gg1$, whereas it approaches infinity in the non-relativistic limit $\sigma\ll1$.

This fit is only applicable to the slopes derived from the main component of the spectrum, i.e., we exclude the additional high-energy component found for $\beta\sim\beta_{\rm max}$. In addition, the steepness of the spectra for $\beta \gtrsim 0.1$ limits the robustness of the fits beyond this $\beta$ value . For this reason, the fits above $\beta\sim0.1$ are indicated with dotted lines.

In addition to the power-law slope, we have also quantified the efficiency of non-thermal electron acceleration on $\beta$ and $\sigma$ of the plasma.  It is evident from Figure \ref{efficiency_fit} that the dependence of the efficiency on $\sigma$ and $\beta$ mirrors the trends described above for the power-law slope. At low $\beta$, where the power-law slope is hard, the efficiency saturates at a value that is independent of $\beta$ but is systematically larger for higher values of $\sigma$. In the other extreme, for $\beta \gtrsim 3 \times 10^{-3}$, because the electron spectrum becomes significantly softer, the non-thermal efficiency approaches zero.

The combined dependence of the electron non-thermal efficiency $\epsilon$ on plasma $\beta$ and magnetization $\sigma$ can be empirically fit as 
\begin{equation}
\epsilon = A_{\epsilon} + B_{\epsilon} \tanh{(C_{\epsilon}\beta)}~,
\label{fiteff}
\end{equation}
where
\begin{equation}
\begin{aligned}
A_{\epsilon}=1 - \frac{1}{4.2 \sigma^{0.55}+1}~,~B_{\epsilon}=0.64\,\sigma^{0.07}~,~C_{\epsilon}=-68\,\sigma^{0.13}   ~.   
\end{aligned}
\end{equation}
We show the fits in Figure \ref{efficiency_fit} with solid lines. In our empirical fit, the efficiency tends to zero for $\sigma\ll1$ (i.e., in the limit of non-relativistic reconnection) and towards 1 for $\sigma\gg 1$ (in the limit of ultra-relativistic reconnection).

\section{Electron Acceleration Mechanisms}\label{mechanism}
In order to understand the dependence of the electron spectrum on $\beta$, it is instructive to investigate the physics of electron acceleration in our simulations.  To this end, we follow individual trajectories of the highest energy electrons in order to identify where they gain most of their energy and what the physical processes responsible for their acceleration are. We focus here on a few representative high-energy electrons. In a forthcoming paper, we will explore the physics of electron acceleration in greater detail.

\begin{figure*}[!t]
\centering
\includegraphics[width =.7\textwidth]{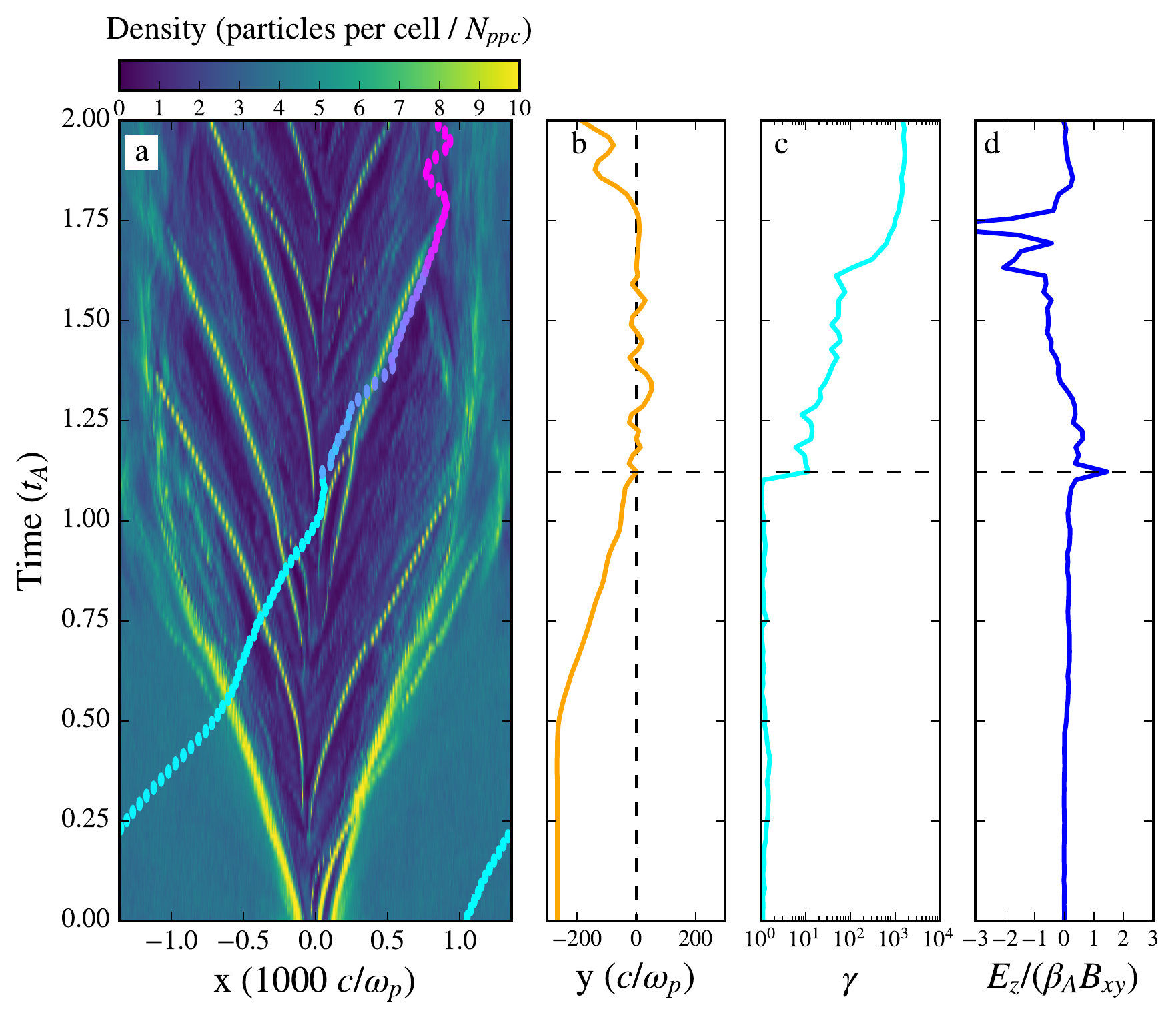}
\caption{Representative electron trajectory from a simulation with $\sigma=0.3$ and $\beta=3\times 10^{-4}$ (simulation B1), whose temporal evolution of particle density and energy spectra is presented in  Figures \ref{lowbeta_threeplot} and \ref{timespec}, respectively. The vertical axis represents time in units of the Alfv\'en crossing time $t_A$.
The background color in panel (a) shows the space-time diagram of  particle density, composed of a sequence of 1D slices taken at $y=0$ (i.e., along the plane of the current sheet). The temporal evolution of the particle $x$-location is overplotted with points, whose color corresponds to the electron energy (from cyan at the initial time, to pink at the final time). In panel (b), the orange line presents the time evolution of the particle $y$-position. Its first interaction with the current sheet is marked with the dashed horizontal line. In panel (c), we show the electron Lorentz factor $\gamma$.  In panel (d), we plot the temporal evolution of the quantity $E_{z}/\beta_{A}B_{xy}$ measured at the particle location, which proves to be a useful diagnostic of the particle acceleration mechanisms. We find that the electron (and in general, all the high-energy electrons in low-$\beta$ runs) is accelerated by non-ideal electric fields at X-points, either in the main layer, or in current sheets formed during plasmoid mergers.}
\label{lowbeta_prtl}
\end{figure*}

\begin{figure*}[!t]
\centering
\includegraphics[width =.7\textwidth]{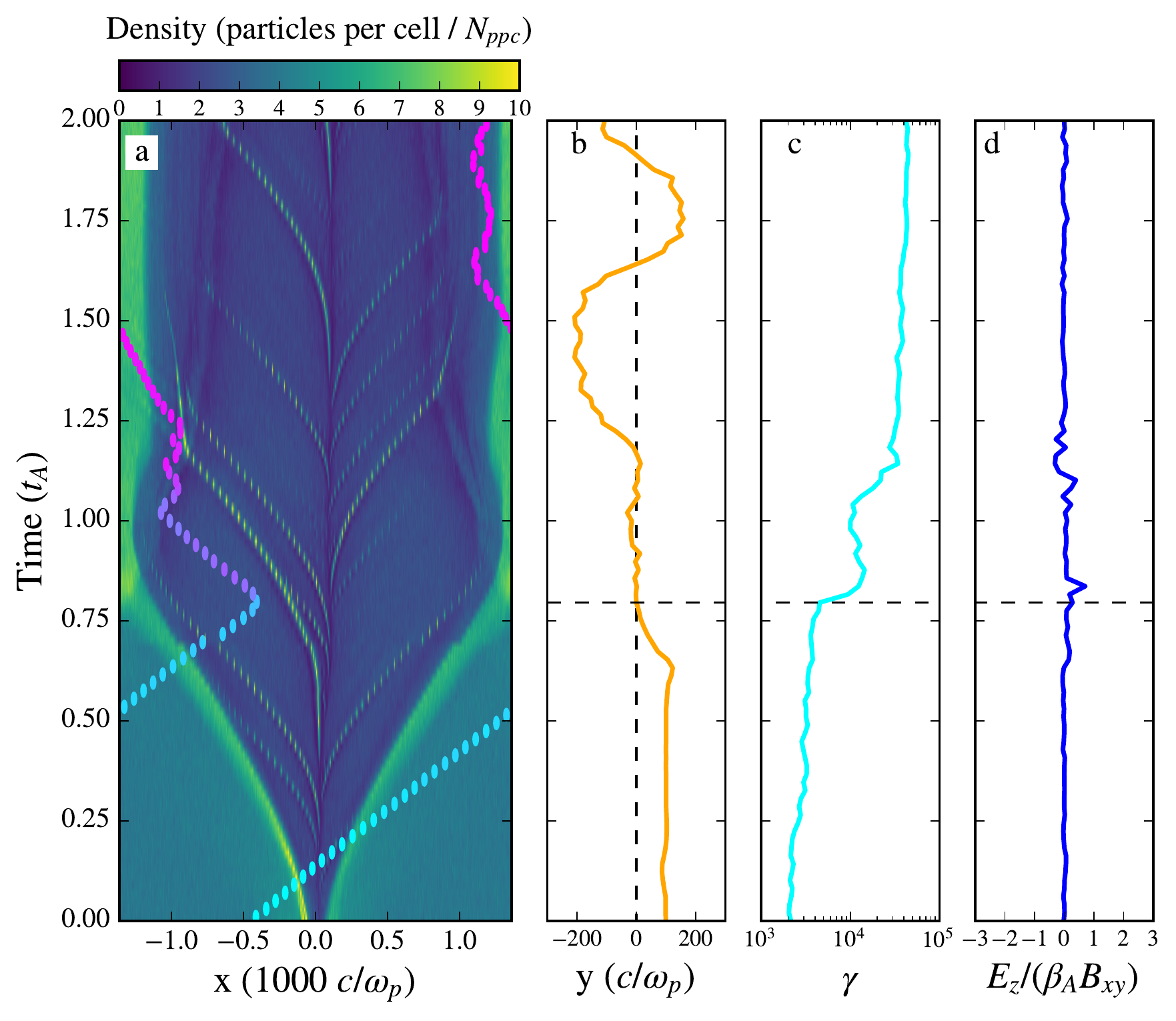}
\caption{Electron trajectory from a simulation with $\sigma=1$ and $\beta=0.16$ (simulation C7), as a representative case of particles (both electrons and protons) belonging to the additional high-energy component appearing for $\beta\approx \beta_{\rm max}$. The temporal evolution of the corresponding electron energy spectrum is shown in Figure \ref{sig1_timespec}. See Figure \ref{lowbeta_prtl} for a description of the content of the panels.  We find that most of the particle energy gain comes from a Fermi-like process, while the electron is bouncing between the reconnection outflow and the edge of the boundary island.}
\label{highbeta_prtl}
\end{figure*}

We show in Figures \ref{lowbeta_prtl} and \ref{highbeta_prtl} examples of representative trajectories of electrons accelerated in low- and high-$\beta$ simulations, respectively.  The electron in the low-$\beta$ case belongs to the main component of particles accelerated by reconnection, whereas the electron in the high-$\beta$ case belongs to the additional spectral component that emerges at $\beta \sim \beta_{\rm{max}}$.  In each panel, the vertical axis represents time in units of the Alfv\'en crossing time $t_A$.  In panel (a), the background color shows a 1D slice of the density, taken along the plane of the current sheet.  The temporal evolution of the  $x$-location of the particle is shown with a sequence of points, with the color corresponding to the particle's energy, starting with cyan and evolving towards pink.  In panel (b), the orange line presents the time evolution of the $y$-position of the particle. Its first interaction with the current sheet (i.e., at $y=0$) is marked with the dashed horizontal line. Note that the $x$-position of the particle depicted in panel (a) can be meaningfully compared with the background density only when the particle is close to the $y=0$ plane, where the density slices in panel (a) are taken.
In panel (c), we show the electron Lorentz factor $\gamma$. 
 In panel (d), we plot the quantity $E_{z}/\beta_{A}B_{xy}$ measured at the particle location, i.e., the out-of-plane electric field $E_z$ divided by the in-plane magnetic field $B_{xy}=(B_{x}^{2}+B_{y}^{2})^{1/2}$ and by the dimensionless Alfv\'en velocity $\beta_A=\sqrt{\sigma/(1+\sigma)}$. This will prove to be a useful diagnostic of the particle acceleration mechanisms, for the following reason: reconnection outflows move at roughly the Alfv\'en speed, so the electric fields carried by a magnetic field $B_{xy}$ are expected to be $E_{z,\rm ideal}\sim \beta_{A}B_{xy}$, in ideal MHD. On the other hand, in regions of strong magnetic dissipation  (e.g., at  X-points), non-ideal electric fields can largely exceed the MHD expectation. Because of that, when the ratio $E_{z}/\beta_{A}B_{xy}$ exceeds unity, it is likely that the particle is experiencing a strong non-ideal electric field, which can serve as an efficient particle accelerator.

\subsection{Electron Acceleration at Low $\beta$}\label{mechanisms}
We show in Figure \ref{lowbeta_prtl} a representative high-energy electron extracted from a simulation with $\sigma=0.3$ and $\beta=3\times 10^{-4}$. For this case, we have presented the temporal evolution of the particle density in Figure \ref{lowbeta_threeplot} and of the electron and proton energy spectra in Figure \ref{timespec}.

A comparison of panels (b) and (c) demonstrates that the electron is first accelerated when it interacts with the current sheet for the first time. During this first interaction with the layer, the particle experiences a value of $E_{z}/\beta_{A}B_{xy}$ larger than unity (see panel (d)), indicating that the acceleration is driven by non-ideal electric fields. In fact, panel (a) shows that during this acceleration episode, the electron is located in one of the under-dense regions associated with X-points. Accelerated by the non-ideal electric field, the electron Lorentz factor at the X-point quickly increases from $\gamma\approx 1$ up to $\gamma \approx 20$.

The electron is then trapped in a secondary plasmoid, which can be identified in panel (a) as the yellow structure that the particle orbit follows at $1.2\lesssim t/ t_A\lesssim 1.7$. While in the plasmoid, the electron energy stays nearly constant, aside from a moderate increase (by roughly a factor of two) when the electron moves from the trailing to the leading edge of the plasmoid at $t\simeq 1.3\, t_A$. 

At $t\simeq 1.7\, t_A$, when the plasmoid merges with the boundary island, the electron lies in between the two. At the interface of the two  merging structures, a current sheet forms along the $y$ direction, i.e., perpendicular to the main reconnection layer (e.g., see the interface at $x\approx -1500 \, c/\omega_{p}$ in Fig. \ref{lowbeta_threeplot}c). As it happens for the main layer, the newly developed current sheet breaks into a series of secondary plasmoids separated by X-points. At one of such X-points, the non-ideal electric field further increases the electron energy up to  $\gamma \approx 10^3$. The role of the non-ideal electric field in this episode of acceleration is evident in panel (d), where $E_{z}/\beta_{A}B_{xy}$ peaks sharply at $t\simeq 1.7\, t_A$. Its negative sign is consistent with the fact that the non-ideal electric field in between merging plasmoids is expected to have opposite direction than in the main layer. 

While many low-$\beta$ electron trajectories resemble the one we have presented here, some electrons show only one episode of acceleration, analogous to either the first or the second stage shown in Figure \ref{lowbeta_prtl}. In other words, 
some electrons pick up all of their energy at an X-point during their first interaction with the current sheet, while others are accelerated at current sheets formed when secondary plasmoids merge with each other or with the boundary island.  In either case, in low-$\beta$ simulations, all the high-energy electrons are predominantly accelerated by non-ideal electric fields associated with reconnecting magnetic fields, either at the primary X-point, at secondary X-points, or in current sheets formed during plasmoid mergers.

\subsection{Electron Acceleration at $\beta\approx \beta_{\rm max}$}
We show in Figure \ref{highbeta_prtl} the trajectory of a representative electron from a simulation with $\sigma=1$ and $\beta=0.16$. This is the simulation that shows the strongest signature of the additional high-energy component appearing at late times for $\beta\approx \beta_{\rm max}$. The temporal evolution of the corresponding electron spectrum is shown in Figure \ref{sig1_timespec}.

Two phases of energization are seen in the time evolution of the electron energy in panel (c). The first episode, when the electron's Lorentz factor increases from $\gamma \approx 2\times 10^3$ up to $\gamma \approx 10^4$, is associated with the first encounter with the current sheet. However, panel (a) shows that the electron here interacts with the unstructured outflow, and not with an X-point as in the low-$\beta$ case. As a result, the value of $|E_{z}|/\beta_{A}B_{xy}$ along the electron trajectory is much smaller than in the low-$\beta$ case. In fact, most of the inflowing electrons in this simulation experience this acceleration episode at their first encounter with the current sheet, regardless of where they interact.  Since such an energization phase is common to the majority of electrons, it should be regarded as bulk heating, rather than non-thermal particle acceleration. Indeed, an electron with $\gamma \approx 10^4$ (as appropriate for the electron in Figure \ref{highbeta_prtl}, after the first energization episode) would not belong to the high-energy spectral component seen in Figure \ref{sig1_timespec} (which lies at $\gamma\gtrsim 2\times 10^4$). 

In its second energization episode, the electron is accelerated up to $\gamma \approx 7\times 10^4$ after it reaches the outskirts of the boundary island at $t\simeq t_A$.  At this point, its energy is within the energy range covered by the high-energy component in Figure \ref{sig1_timespec}.  From $t\simeq t_A$ to $t\simeq1.2\, t_A$, it stays confined between the boundary island and the reconnection outflow. We attribute the energy increase in this phase to a Fermi-type process in between converging flows (i.e., the reconnection outflow and the boundary island), for two main reasons: (\textit{i}) as in the first phase of energization, this second episode does not arise from a strong non-ideal electric field, which would be expected for X-point acceleration; (\textit{ii}) the fractional energy gain is comparable between the first and second phases of energization, as expected for a Fermi-like process (see the next subsection).

We find that all of the highest energy electrons in $\beta\approx \beta_{\rm max}$ simulations show this Fermi-type acceleration as they get trapped between the reconnection outflow and the boundary island.  The highest energy protons in  $\beta\approx \beta_{\rm max}$ simulations also display the same acceleration physics as electrons, which explains the similarity between the energy spectra of the two species (compare red and cyan lines in Figure \ref{sig1_timespec}).

\subsection{Comparing the Acceleration Mechanisms}
In this subsection, we present a few qualitative arguments to justify why X-point acceleration plays a more significant role at low $\beta$, whereas the Fermi process is predominant at high $\beta$ (and more specifically, at $\beta\approx \beta_{\rm max}$). We defer a more detailed analysis of the physics of particle acceleration to a future study.

First, as shown in Figures \ref{sig_3_twoplot} and \ref{sig3_twoplot}, low-$\beta$ simulations display a much higher number of secondary plasmoids, and consequently, of secondary X-points, than high-$\beta$ runs. It follows that the fraction of inflowing electrons that are likely to enter the current sheet at the location of an X-point --- where they can be accelerated by non-ideal electric fields --- is higher at lower $\beta$, resulting in higher acceleration efficiencies.

Second, the strength of the reconnection electric field $E_z$ is proportional to the particle inflow rate (i.e., to the reconnection rate), which steadily decreases as $\beta$ increases, as shown in Figure \ref{inflow_rates_over_Va}. So, the non-ideal electric field will be weaker at higher $\beta$, resulting in a slower rate of particle acceleration at X-points.

Finally, we can compare the typical energy gains expected from one episode of X-point acceleration and one Fermi cycle, as a function of $\sigma$ and $\beta$. The electron energy gain at an X-point will be equal to the work performed by the non-ideal electric field. Setting this to be $\sim 0.1\,\beta_{A}$ of the upstream magnetic field $B_0$, we get
\begin{equation}
\Delta\gamma_{e,\rm X} m_{e} c^{2} \approx  0.1 \beta_{A} e B_{0} L~,
\end{equation}
where $L$ is the length of the acceleration region in the $z$-direction. If $L$ is normalized to the proton skin depth, with $L=L_{di} \,c/\omega_{pi}$, we find
\begin{equation}\label{xpoint-gain-temp}
\Delta \gamma_{e,\rm X} \approx  0.1 \frac{m_{i}}{m_{e}} \frac{\sigma}{\sqrt{\sigma+1}} L_{di}~.
\end{equation}
Clearly, the energy gain for X-point acceleration is insensitive to the initial electron temperature. On the other hand, the fractional energy increase per Fermi cycle is $\sim \beta_A$, if particles bounce between the reconnection outflow, which is moving at $\sim v_A$, and the boundary island, which is stationary.\footnote{We are also implicitly assuming that the converging flows are non-relativistic, which requires $\sigma\lesssim 1$ (so that the Alfv\'en speed is non-relativistic).} It follows that
\begin{equation}\label{fermi}
\Delta \gamma_{e,\rm Fermi}\approx \beta_A \theta_e~,
\end{equation}
and if protons and electrons are set up in temperature equilibrium, this becomes
\begin{equation}\label{fermi}
\Delta \gamma_{e,\rm Fermi}\approx \beta \frac{m_i}{m_e} \frac{\sigma^{3/2}}{\sqrt{\sigma+1}}~.
\end{equation}

This simple argument shows that, for fixed $\sigma$, X-point acceleration will provide a larger energy gain at low $\beta$, whereas the Fermi process will be energetically dominant in the high-$\beta$ regime.

\section{Conclusions}\label{conclusions}
In this work, we have investigated with large-scale 2D PIC simulations the physics of non-thermal particle acceleration in trans-relativistic  reconnection, covering a very wide parameter space in $\sigma$ and $\beta$ and employing the physical proton-to-electron mass ratio. For four values of the magnetization ($\sigma=$ 0.1, 0.3, 1, and 3), we have explored a wide range of $\beta$, from $\beta=10^{-4}$ up to the maximum possible value of $\beta$, that is $\beta_{\rm max}\approx 1/4\sigma$. 

We find that the electron spectrum in the reconnection region can be generally modeled as a non-thermal power law, but the properties of the spectrum are strongly dependent on $\beta$. At $\beta \lesssim 3 \times 10^{-3}$, electron acceleration is efficient and the electron spectrum is dominated by a hard power law.  Its slope is insensitive to $\beta$ and depends on $\sigma$ as $p\simeq 1.8 +0.7/\sqrt{\sigma}$, in agreement with the result by \citet{werner2018} (who considered a single value of $\beta=0.01$).  By tracking a large number of particles in our simulations, we find that that in this low-$\beta$ regime, electrons are primarily accelerated by the non-ideal electric field at X-points, either  in the initial current layer  or in current sheets generated in between merging magnetic islands.

At higher $\beta$, the electron power law steepens significantly, and the electron spectrum eventually approaches a Maxwellian distribution, for all values of $\sigma$.  In other words, the efficiency of non-thermal electron acceleration approaches zero. At high values of $\beta$ near $\beta_{\rm max}\approx1/4\sigma$, when both electrons and protons start relativistically hot, the spectra of both species display an additional component at high energies, containing a few percent of particles, which are accelerated via a Fermi-like process by bouncing in between the reconnection outflow and the stationary magnetic island at the boundary of our periodic domain.

For the main population of non-thermal electrons (i.e., excluding the additional component emerging at $\beta\rightarrow \beta_{\rm max}$),
we provide an empirical prescription for the dependence of the power-law slope and the acceleration efficiency on $\beta$ and $\sigma$. We also measure the inflow rate (i.e., the reconnection rate) as a function of $\beta$ and $\sigma$, and find that, for a given $\sigma$, the reconnection rate steadily decreases with increasing $\beta$.

Our results can provide a physically-grounded prescription for non-thermal electron acceleration via magnetic reconnection, in a regime relevant to hot accretion flows like Sgr A* at our Galactic center (e.g., \citealt{ressler2015}; \citealt{ball2016}; \citealt{mao2017}; \citealt{chael2017}). When implemented as subgrid models into global MHD simulations, our findings have the potential to unveil the origin of the flaring behavior of Sgr A* (\citealt{ponti2017}).

We conclude with a few caveats. First, our simulations have employed a 2D setup, and it will be important to see whether 3D effects alter the physics of electron acceleration and the resulting electron energy spectra.  Second, we have only considered reconnection setups with no guide fields and equal electron and proton temperatures.  However, for application to accretion flows around black holes, we generally expect non-zero guide fields in reconnection regions (\citealt{ball2017}) and protons to be significantly hotter than electrons.  
 We will explore these effects in future studies.

\section*{Acknowledgements}
We thank Michael Rowan, Dimitrios Psaltis, Ramesh Narayan, and Dimitrios Giannios for useful discussions.  We gratefully acknowledge support for this work from NSF AST-1715061 and Chandra Award No. TM6-17006X. FO acknowledges a fellowship
from the John Simon Guggenheim Memorial Foundation
in support of this work.  LS acknowledges support from DoE DE-SC0016542, NASA Fermi NNX-16AR75G, NASA ATP NNX-17AG21G, NSF ACI-1657507, and NSF AST1716567.  The simulations were performed on El Gato at the University of Arizona, Habanero at Columbia, and NASA High-End Computing (HEC) program through the NASA Advanced Supercomputing (NAS) Division at AMES Research Center.

\appendix

\section{Appendix A: Effects of Box Size}\label{boxsize}
\begin{figure}[!h]
	\centering
	\includegraphics[width =0.5\textwidth]{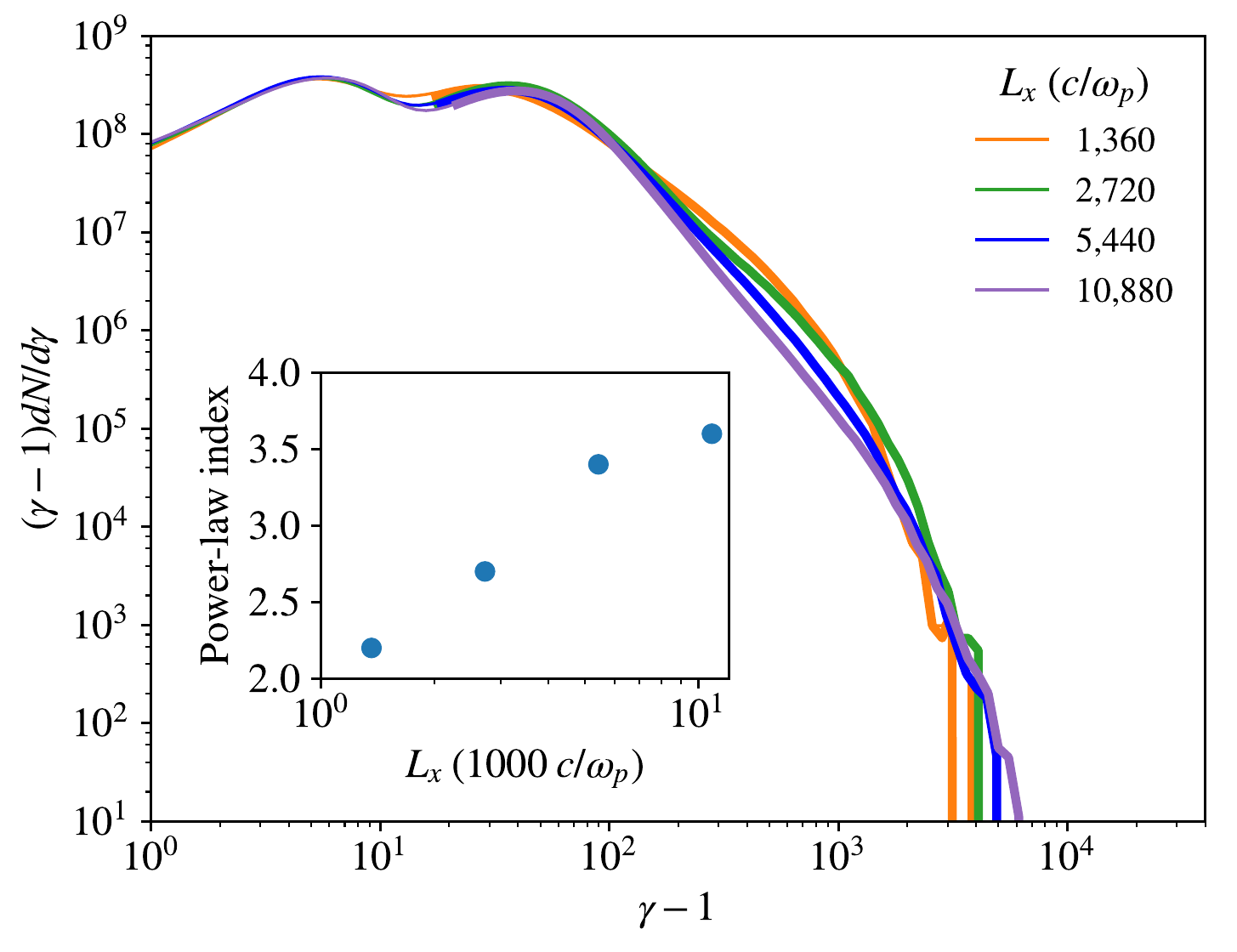}
	\caption{Electron spectra at $t\approx 2\,t_{A}$ from simulations with fixed $\sigma=0.3$ and $\beta=0.006$ but having different box sizes, as indicated in the legend. Inset: electron power-law index as a function of box size. We find that as the box length increases, the power-law index gets larger (so, the non-thermal tail steepens).}
	\label{sigpoint3_boxsize}
\end{figure}

\begin{figure}[t]
	\centering
	\includegraphics[width =0.5\textwidth]{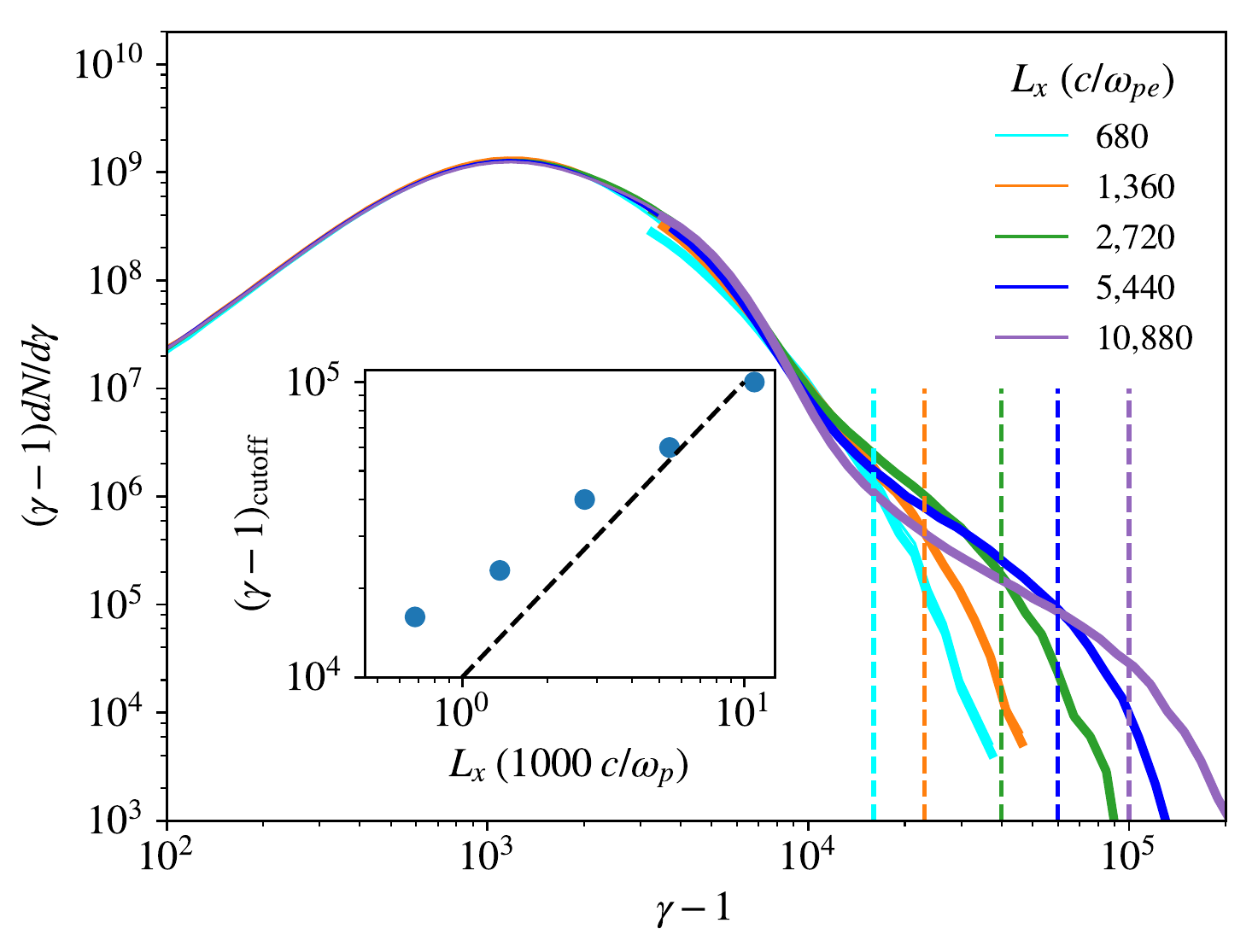}
	\caption{Electron spectra at $t\approx 2\,t_{A}$ from  simulations with fixed $\sigma=1$ and $\beta=0.16$ but having different box sizes, as indicated in the legend. This is a representative case with $\beta\approx\beta_{\rm max}$, which for sufficiently large boxes shows an additional high-energy spectral component. Inset: dependence of the high-energy cutoff (indicated in the main plot by the vertical dashed lines) on domain size, which shows that for large boxes the high-energy cutoff scales nearly linearly with box size (the linear scaling is indicated by the dashed black line).}
	\label{sig1_boxsize}
\end{figure}

Previous studies (e.g., \citealt{werner2018}) have shown that in a larger computational domain, the electron power law tends to steepen, but it extends to higher energies.
In this Appendix, we investigate the dependence of our results on the size of the computational box, for both a low-beta case ($\sigma=0.3$ and $\beta=0.006$, in  Figure \ref{sigpoint3_boxsize}) and a high-beta case ($\sigma=1$ and $\beta=0.16\approx \beta_{\rm max}$, in Figure \ref{sig1_boxsize}). While in \citet{werner2018} the extent of the computational domain in the direction along the reconnection layer was $L_x=120\,r_{e,\rm hot}$ or smaller, here we explore the dependence on box size up to much larger values: for the low-beta case up to $L_x=1,178\,r_{e,\rm hot}$, and for the high-beta case up to $L_x=5,426\,r_{e,\rm hot}$.\footnote{In Figures \ref{sigpoint3_boxsize} and  \ref{sig1_boxsize}, the legend indicates the box length in units of the electron skin depth $c/\omega_p$, rather than the Larmor radius $r_{e,\rm hot}=\sigma_e m_e c^2/eB_0$ of a relativistic electron with Lorentz factor $\sigma_e=\sigma_i m_i/m_e$, which was the unit of length in \citealt{werner2018}.}

In Figure \ref{sigpoint3_boxsize}, we show electron spectra extracted from four simulations with fixed $\sigma=0.3$ and $\beta=0.006$ but having different box sizes, with $L_x/(c/\omega_p)=1,360$, 2,720, 5,440, and 10,880 (corresponding to $L_x/r_{e,\rm hot}=147.2$, 294.5, 589, and 1,178). For easier comparison, the normalization of the spectrum is scaled by $\propto L_x^{-2}$.
We find a systematic trend of steeper slopes of the electron non-thermal tail at larger boxes. In the inset of Figure \ref{sigpoint3_boxsize}, we present the dependence of our best-fit slope on the domain size (notice the log-linear scale). In our two largest boxes, the slope seems to saturate at $p\simeq 3.5$. Because $L_x=5,440\,c/\omega_p$ is the choice employed in the main body of the paper, this gives us confidence that we are capturing the asymptotic properties of the electron non-thermal spectrum. While the slope seems to saturate for the two largest boxes, the high-energy cutoff of the spectrum keeps steadily increasing, albeit at a slower rate than the linear scaling found by \citet{sironi2016} in relativistic pair reconnection. Even larger domains will be required to assess the asymptotic scaling of the high-energy cutoff with domain size.

In Figure \ref{sig1_boxsize}, we investigate the dependence on box size of our results for a suite of five simulations (see the legend) with $\sigma=1$ and $\beta=0.16$, a case that displays the additional high-energy component appearing when $\beta\approx \beta_{\rm max}$. We find that the presence of this additional component can be captured only in large domains and is virtually undetectable in the smaller boxes (see the cyan and orange lines). The normalization of the additional component (i.e., the fraction of particles it contains) is a weak function of domain size, but its high-energy cutoff linearly increases with increasing box size.  This can be seen in the inset by comparing the data points for $L_x\gtrsim 3,000\,c/\omega_{p}$ with the linear scaling of the black dashed line. This emphasizes once more the importance of large simulation domains in unveiling the physics of trans-relativistic magnetic reconnection.

\section{Appendix B: Effects of Boundary Conditions and Initial Setups}\label{untriggered}
In the simulations presented in the main body of the paper, we trigger reconnection at the center of our computational domain and we employ periodic boundary conditions in the direction of the reconnection outflows (see Section \ref{setup}). In this Appendix, we explore the effect of different choices of boundary conditions (outflow vs periodic) and initial setups (untriggered vs triggered). For the untriggered runs presented here,  we use periodic boundaries and employ an initial current sheet that is thinner than for our triggered runs ($\Delta=20\,c/\omega_p$, as compared to $\Delta=80\,c/\omega_p$ in our triggered runs), in order to allow the primary tearing mode to develop quickly and to produce several primary X-points in our simulation domain. For the simulations with outflow boundary conditions (as described in \citealt{sironi2016}), we employ a triggered setup with $\Delta=80\,c/\omega_p$.

\begin{figure}[!t]
	\centering
	\includegraphics[width =0.5\textwidth]{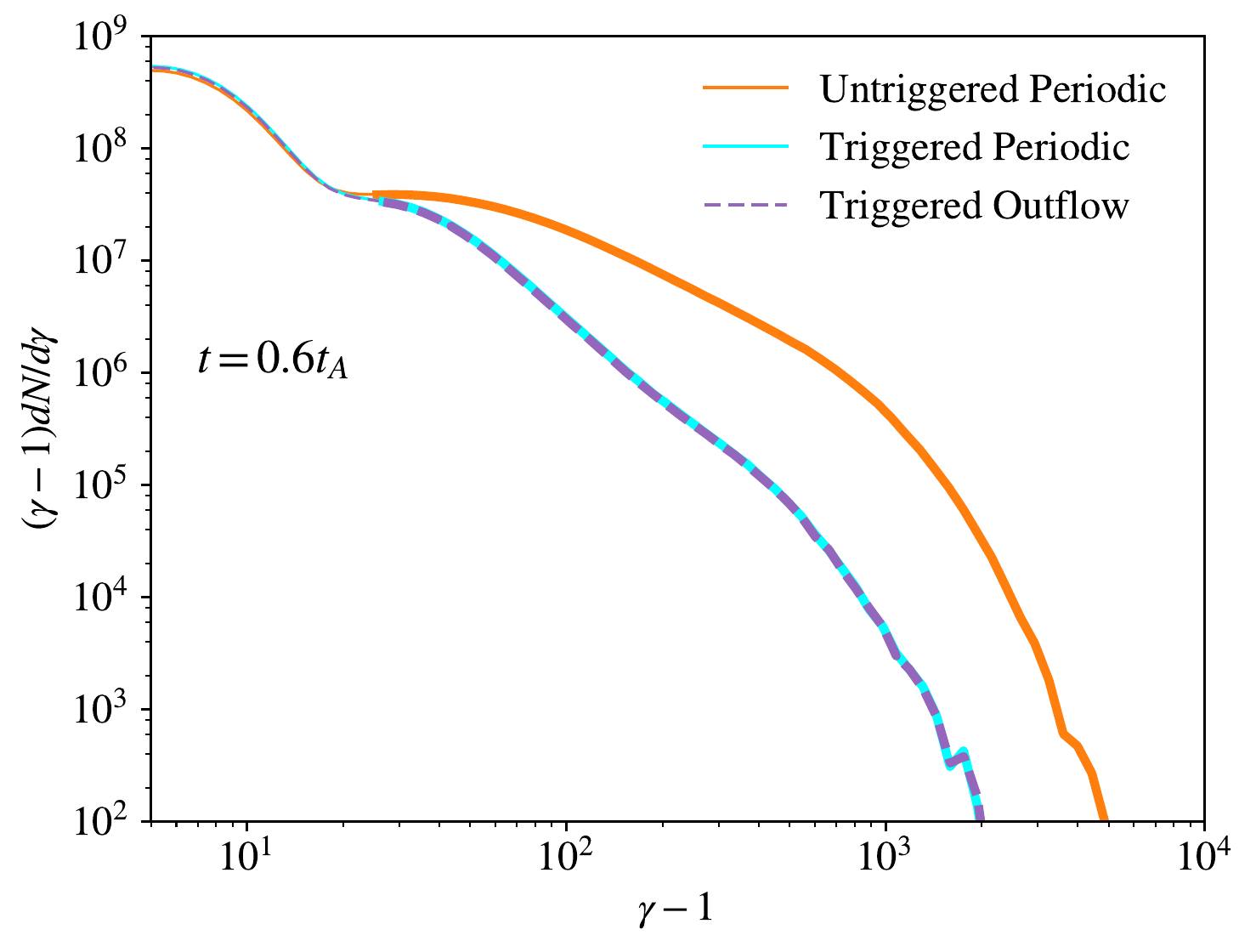}
	\includegraphics[width =0.5\textwidth]{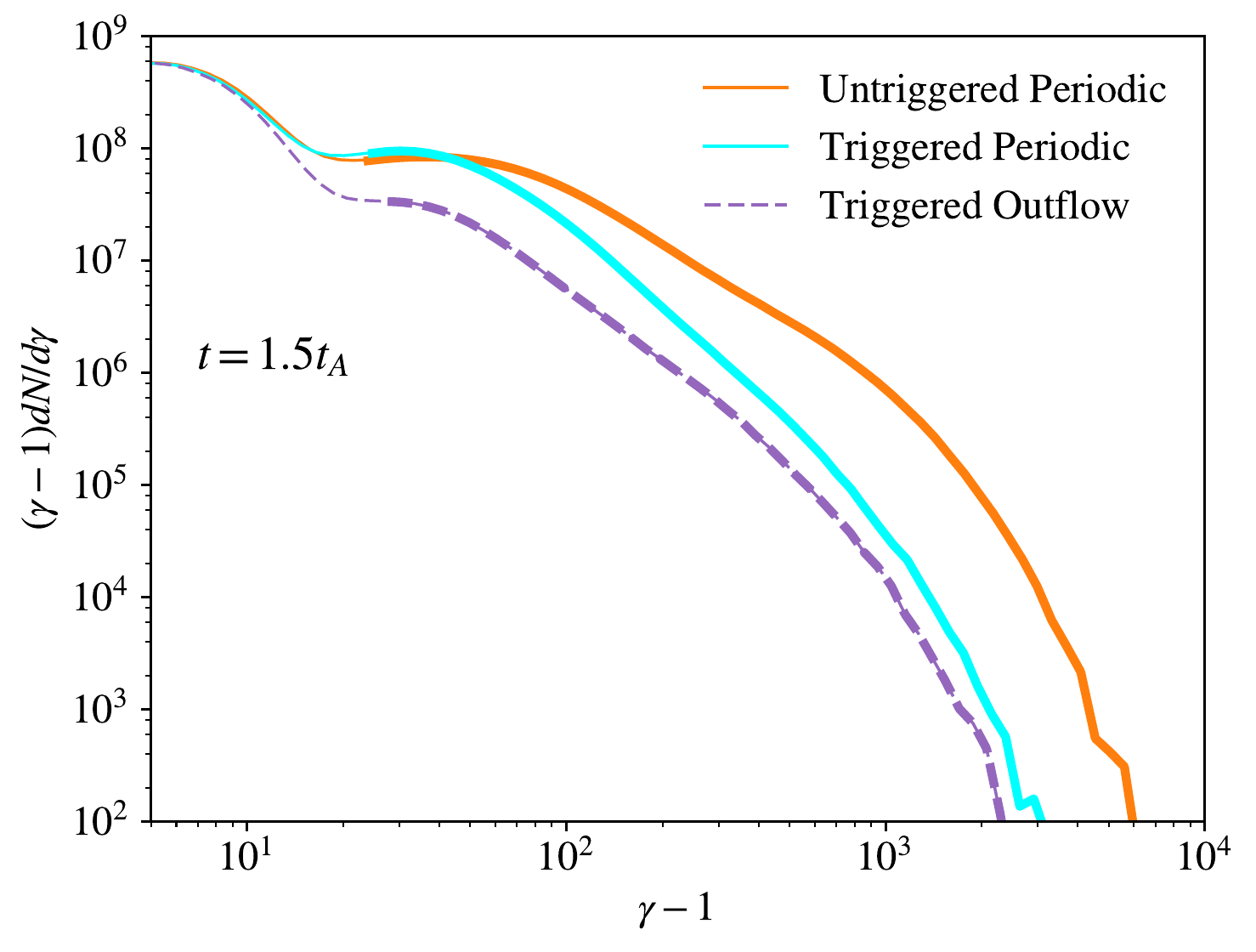}
	\caption{Electron spectra extracted from simulations with $\sigma=0.3$ and $\beta=0.006$ at 0.6 (top) and 1.5 (bottom) Alfv\'en crossing times.  We investigate the dependence on boundary conditions and initial setups. Our fiducial triggered-periodic simulation is shown in cyan, the triggered-outflow case in dashed purple, and the untriggered-periodic in orange. We generally find that the untriggered setup gives harder electron spectra.}
	\label{sigpoint3_outflow_early}
\end{figure}

\begin{figure}[!t]
	\centering
	\includegraphics[width =0.5\textwidth]{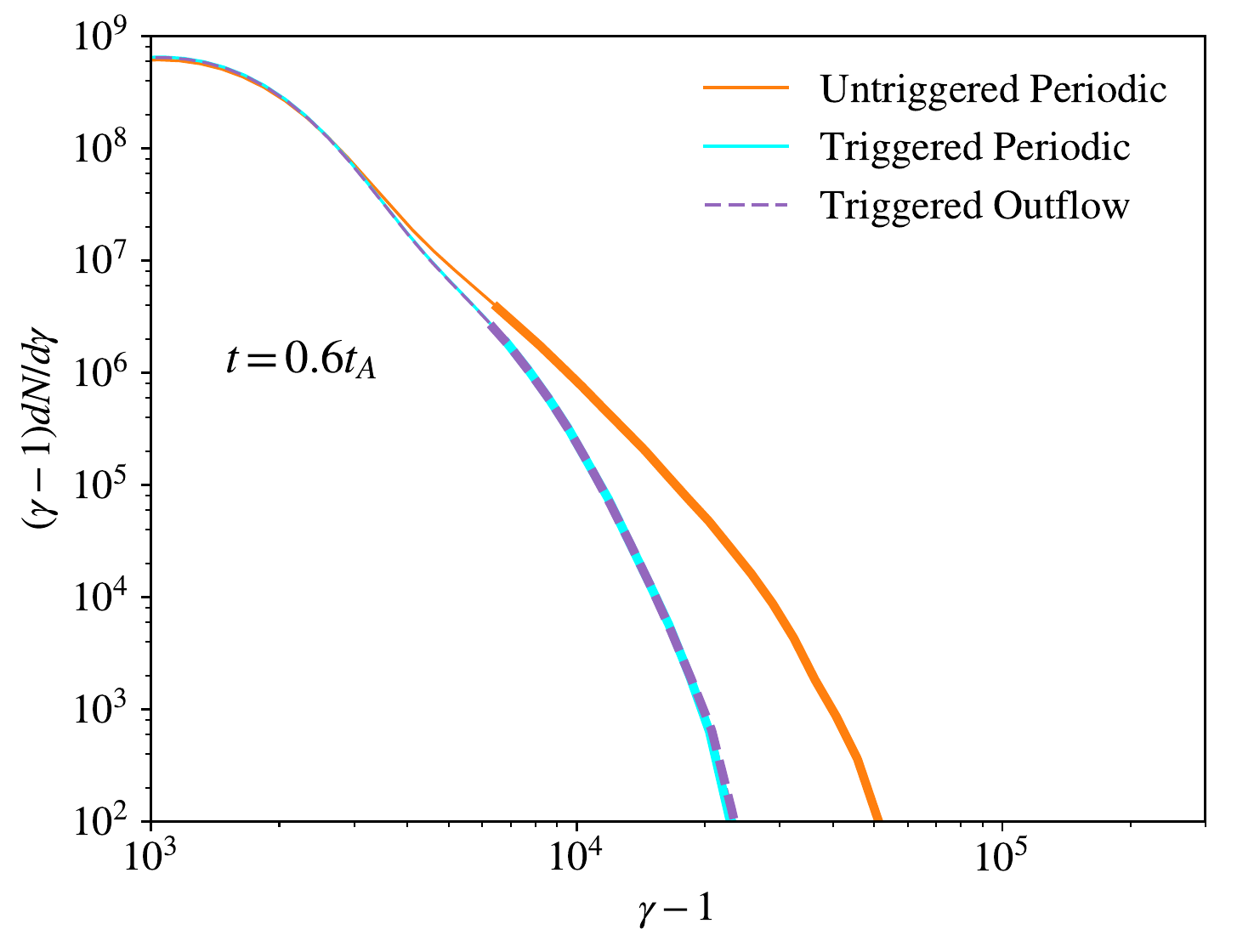}
	\includegraphics[width =0.5\textwidth]{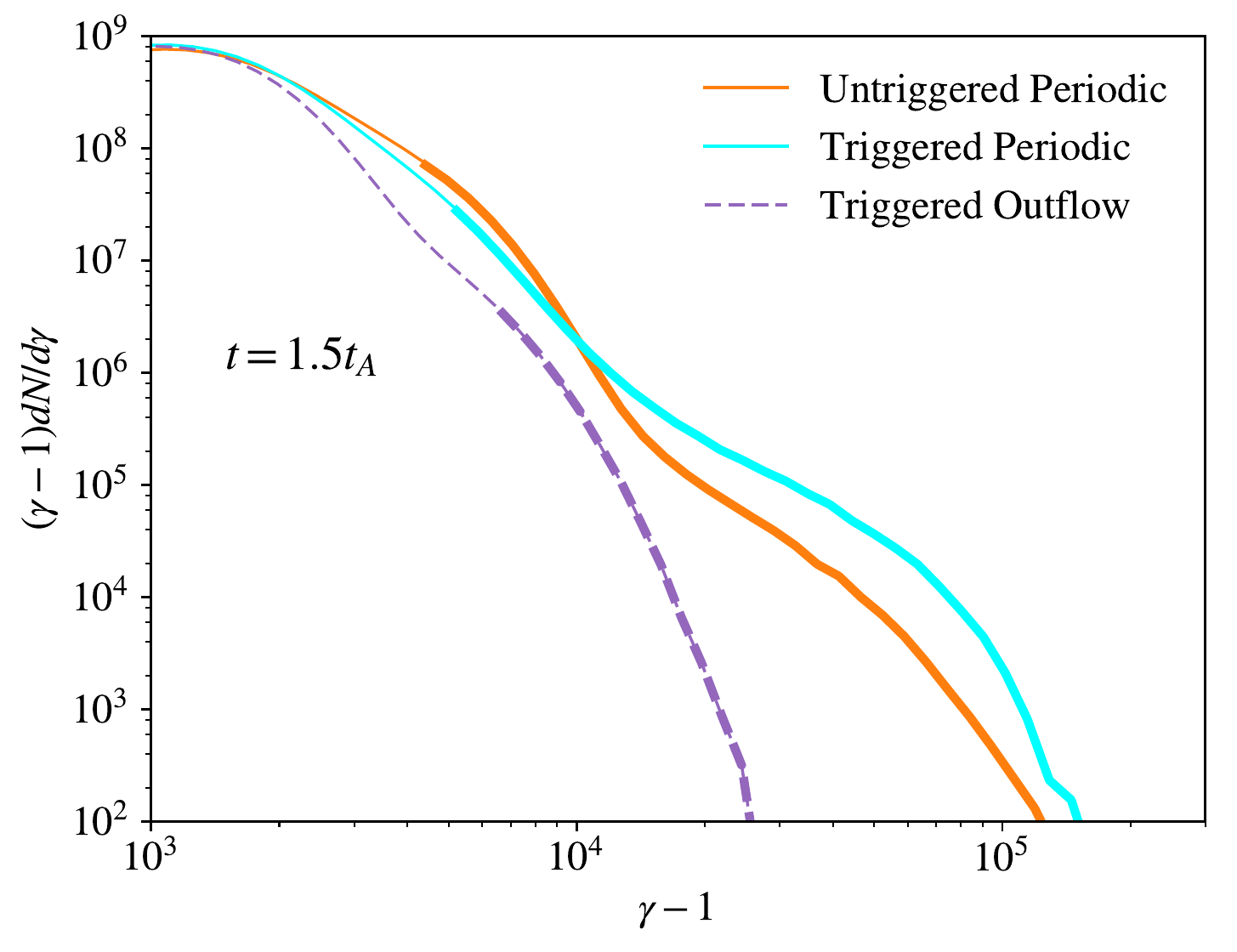}
	\caption{Electron spectra extracted from simulations with $\sigma=1$ and $\beta=0.16$ at 0.6 (top) and 1.5 (bottom) Alfv\'en crossing times.  We investigate the dependence on boundary conditions and initial setups. Our fiducial triggered-periodic simulation is shown in cyan, the triggered-outflow case in dashed purple, and the untriggered-periodic in orange. We find that the additional high-energy component appearing in cases with $\beta\approx \beta_{\rm max}$ is present in both triggered and untriggered setups, at sufficiently late times.}
	\label{sig1_outflow_early}
\end{figure}

We show in Figures \ref{sigpoint3_outflow_early} and \ref{sig1_outflow_early} the spectra of simulations with identical physical and numerical parameters, but different choices of boundary conditions and triggering mechanisms. Figure \ref{sigpoint3_outflow_early} shows the low-$\beta$ case with $\sigma=0.3$ and $\beta=0.006$, whereas Figure \ref{sig1_outflow_early} presents a high-beta case with $\sigma=1$ and $\beta=0.16\approx \beta_{\rm max}$. Top panels refer to early times ($t=0.6\,t_A$), whereas bottom panels to later times ($t=1.5\,t_A$). In both Figures, we compare our standard choice of triggered reconnection with periodic boundaries (cyan lines) with the cases of untriggered reconnection with periodic boundaries (orange lines) and of triggered reconnection with outflow boundaries (purple dashed lines).

In the low-$\beta$ case presented in Figure \ref{sigpoint3_outflow_early}, at early times (top panel), the spectra of triggered-outflow and triggered-periodic simulations are identical, because the reconnection fronts have yet to reach the boundaries of the domain and thus, the system is not yet affected by our choice of boundary conditions. The spectrum of the untriggered-periodic case is significantly harder. The difference between the untriggered-periodic case and our standard triggered-periodic choice persists at later times (bottom panel). There, we measure an electron power-law index of 2.7 for the untriggered case, while the triggered simulation has a power-law index of 3.4.

We defer a detailed investigation of the comparison between triggered and untriggered simulations to a later study. Still, we  speculate that the difference may be due to the large number of primary X-points produced by the untriggered setup, which can serve as efficient sites of electron acceleration in low-$\beta$ reconnection (see Section \ref{mechanism}). On the other hand, in our standard choice of a triggered setup, only one primary X-point is formed. If primary X-points are more effective in accelerating electrons than secondary X-points (which are copiously produced in both setups), this can explain the difference in spectral slope. Alternatively (or, additionally), the difference might be attributed to the fact that in untriggered runs, primary islands mostly grow due to ``major'' mergers with other primary islands of comparable size. It follows that the reconnection layer formed in between two merging primary islands is as long as their width. In contrast, in triggered simulations we have only one primary island (the boundary island), and the ``minor'' mergers of secondary plasmoids with the boundary island form shorter layers (whose length is the width of the merging secondary plasmoid). Since reconnection layers in between merging plasmoids play an important role in electron acceleration (see Section \ref{mechanism}), this might explain the observed difference of electron spectral slopes. Our arguments are further supported by the fact that the spectral slope in untriggered runs shows a much weaker dependence on box size than in triggered runs.  In fact, while the number of primary X-points per unit length is constant in untriggered runs, it steadily decreases with box size in triggered runs.

At late times, the run with outflow boundaries (dashed purple line in the bottom panel of Figure \ref{sigpoint3_outflow_early}) has a slightly harder slope than the triggered-periodic case (solid cyan curve). However, the difference is  smaller than the variation between the triggered-periodic and untriggered-periodic cases. 

We show in figure \ref{sig1_outflow_early} several runs with $\sigma=1$ and $\beta=0.16$. As we have discussed in the main body of the paper, a triggered-periodic run with these physical parameters would show the additional high-energy component appearing when  $\beta\approx \beta_{\rm max}$. At late times (bottom panel), we see this component not only in the triggered-periodic setup (cyan line), but also in the untriggered-periodic case (orange curve). Aside from the small difference in the normalization of the additional component, we conclude that its presence is independent of the triggering choice.  However, no such signature is evident in the triggered-outflow simulation (dashed purple line).  This is likely due to the fact that particles belonging to the additional high-energy component are accelerated by a Fermi-like process in between the reconnection outflow (moving at $\sim v_A$) and the stationary boundary island (or, for untriggered runs, one of the primary islands). In outflow simulations, such a strong convergence of flows does not occur. Still, we expect that, for sufficiently large domains, the velocity difference between the unstructured outflow (moving at $\sim v_A$) and a large secondary plasmoid (which is slow, due to being large; see \citealt{sironi2016}) will promote a fraction of electrons into the additional high-energy component.

\section{Appendix C: Tests of Numerical Convergence}\label{convergence}
\begin{figure}[!h]
	\centering
	\includegraphics[width =0.5\textwidth]{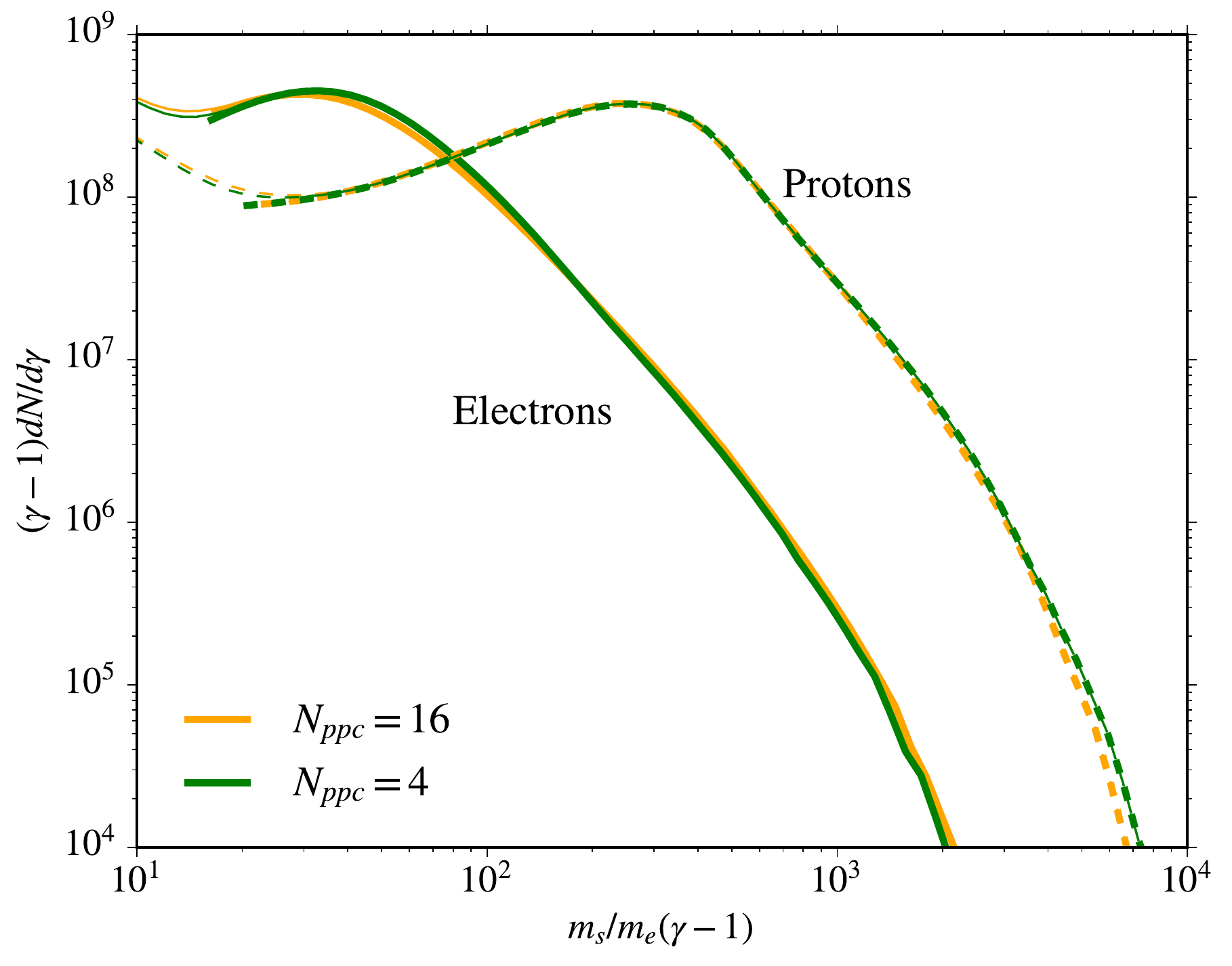}
	\caption{Electron (solid) and proton (dashed) energy spectra for simulations with $\sigma=0.3$ and $\beta=0.006$, where we increase the number of particles per cell from $N_{ppc}=4$  (green) up to $N_{ppc}=16$ (yellow). The spectra are computed at $t\approx 2\,t_{A}$, and are observed to be nearly insensitive to the increase in particles per cell.}
	\label{ppc_spect}
\end{figure}

We have checked that our results are insensitive to the choice of number of computational particles per cell and of spatial resolution. In particular, in Figure \ref{ppc_spect} we compare our results, for a case with $\sigma=0.3$ and $\beta=0.006$, when we increase the number of computational particles per cell from $N_{ppc}=4$ (green curves; solid for electrons, dashed for protons) to $N_{ppc}=16$. Both electron and proton spectra are unchanged.

\begin{figure}[!h]
	\centering
	\includegraphics[width =0.5\textwidth]{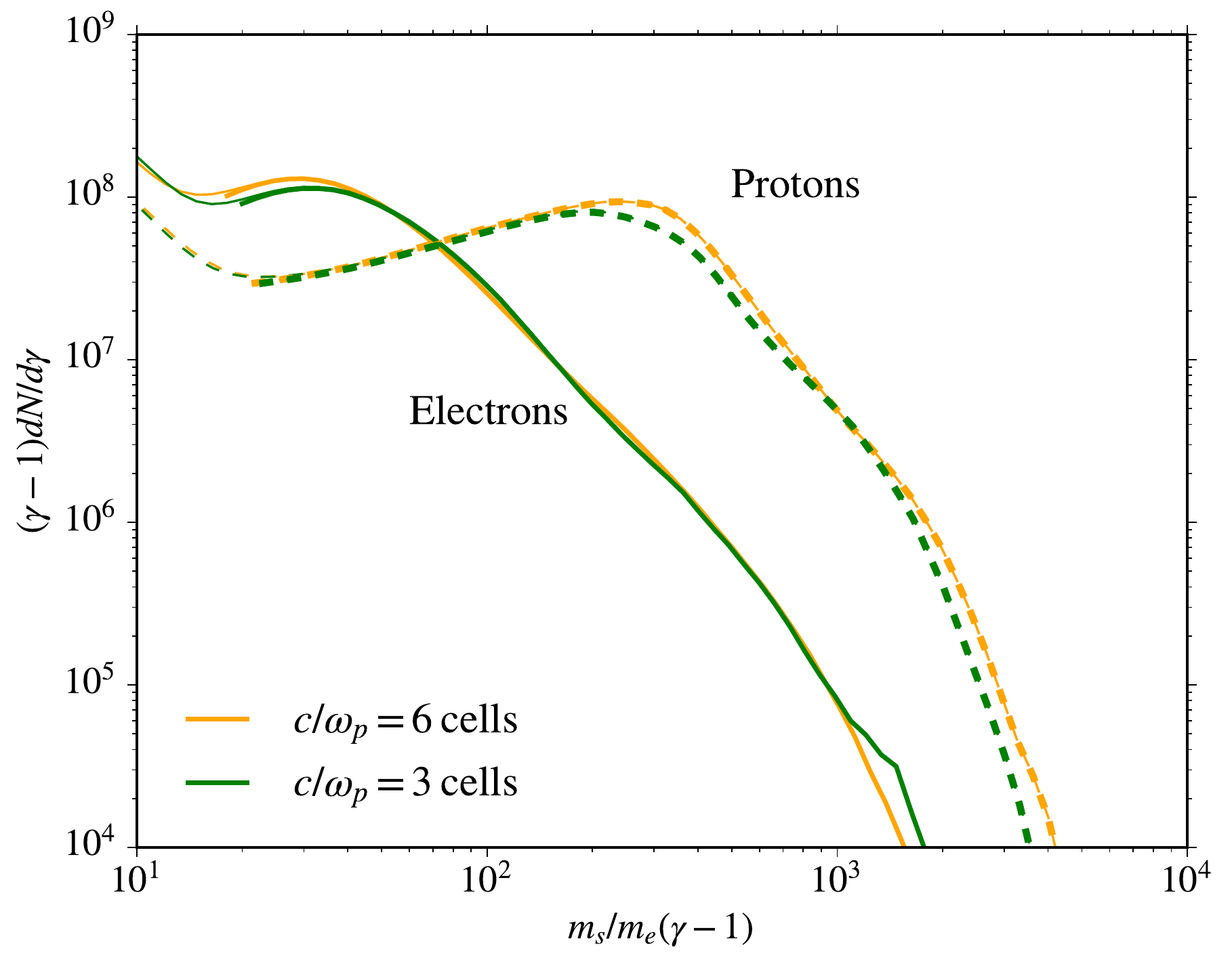}
	\caption{Electron (solid) and proton (dashed) energy spectra for simulations with $\sigma=0.3$ and $\beta=0.006$, where we vary the spatial resolution from $c/\omega_{p}=3$ cells (green) up to $c/\omega_{p}=6$ cells (yellow). The spectra are computed at $t\approx 2\,t_{A}$, and are observed to be nearly insensitive to doubling the resolution.}
	\label{comp_spect}
\end{figure}

In  Figure \ref{comp_spect}, we show the effect of doubling the spatial resolution from $c/\omega_{p}=3$ cells (green) up to $c/\omega_{p}=6$ cells (yellow). In doing so, we increase the number of computational cells along the current sheet and we evolve the simulation for twice as many timesteps, so that our results can be properly compared while having the same value of $L_x/(c/\omega_p)$, and at the same time $t/t_A$. The main features of the electron and proton spectra (and in particular, the slope and high-energy cutoff of the electron spectrum) are the same when doubling the spatial resolution.

\section{Appendix D: Spatial Decomposition of the Electron Spectrum}\label{shells}
\begin{figure}[!h]
	\centering
	\hspace{-0.095in}\includegraphics[width =0.48\textwidth]{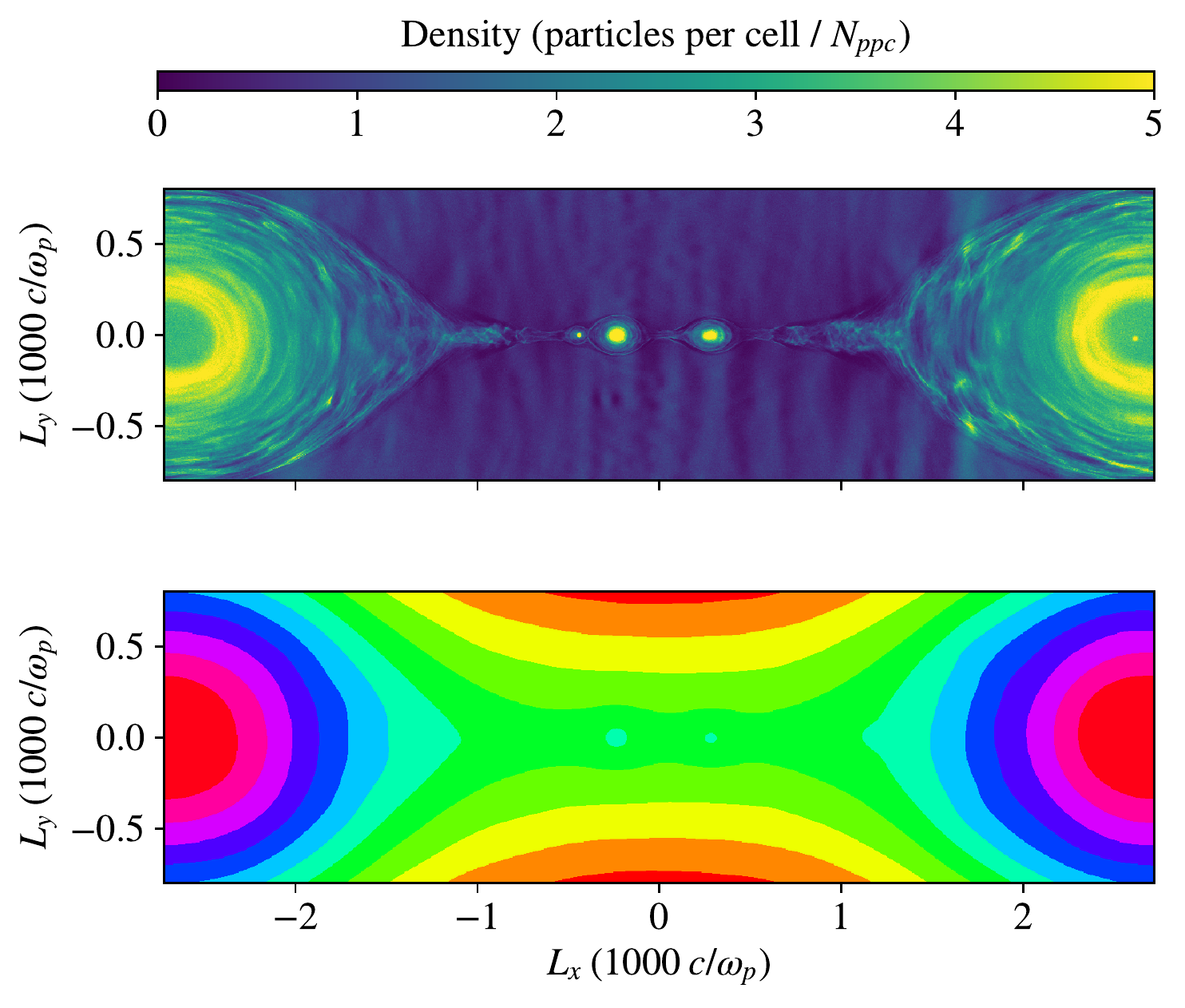}
	\includegraphics[width =0.48\textwidth]{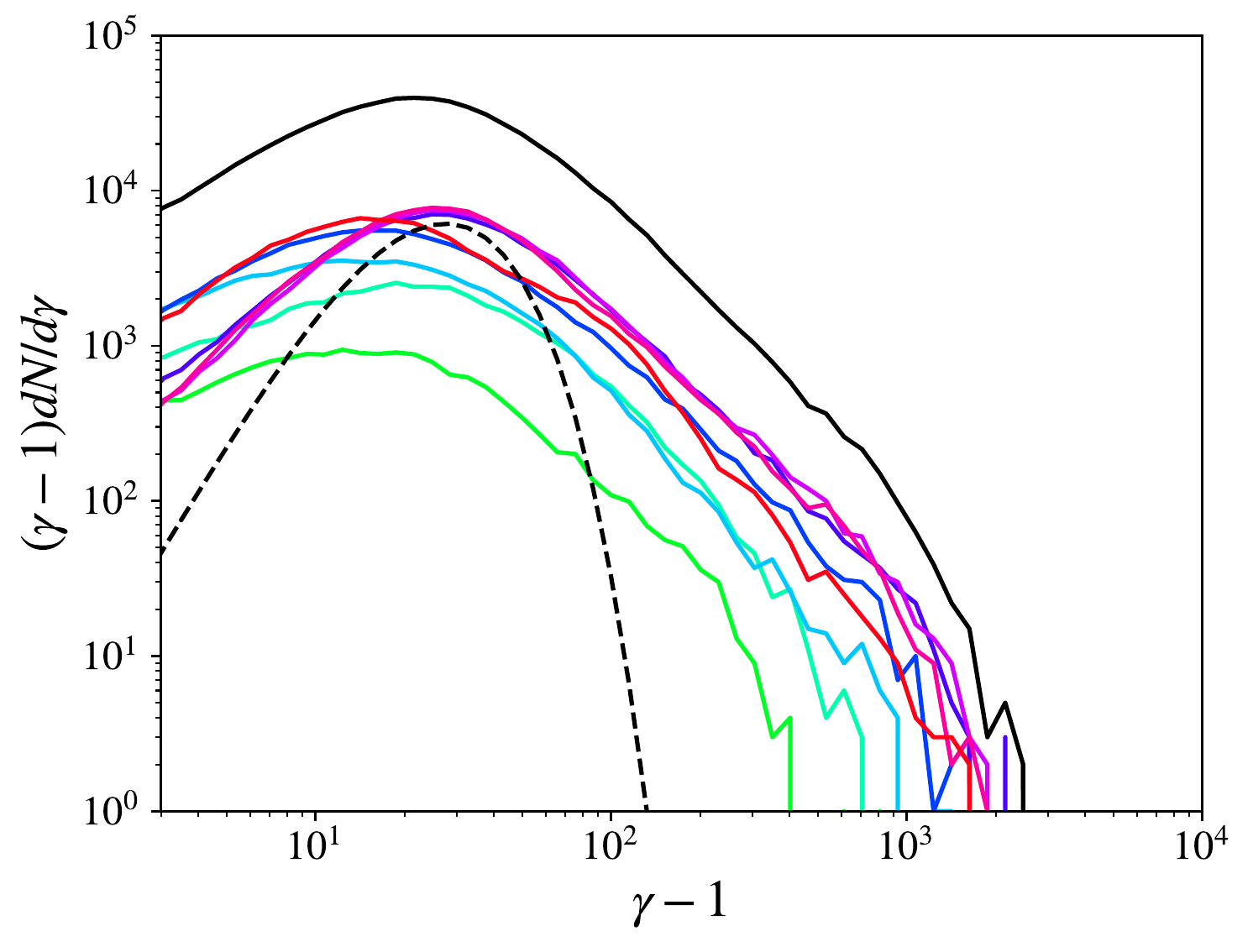}
	\caption{Top: 2D density plot from a simulation with $\sigma=0.3$ and $\beta=3\times10^{-4}$ (simulation B1), taken at $t\approx 2 \,t_{A}$.  Middle: regions bounded by equipotential contours of the $z$-component of the magnetic vector potential.  Bottom: electron spectra calculated in the shells identified in the middle panel (with the same color coding). The solid black line shows the total spectrum and the dashed black line depicts a Maxwellian distribution.}
	\label{shellvec_spect}
\end{figure}

Here, we investigate spatial variations of the electron spectrum within the boundary island at late times, which contains most of the high-energy electrons.  Specifically, we are interested in whether each local energy spectrum is non-thermal.  We consider a simulation with $\sigma=0.3$ and $\beta=3\times10^{-4}$ and calculate the $z$-component of the magnetic vector potential.  We then decompose the boundary island into shells delimited by equipotential contours, with a procedure similar to \citet{li2017} (see the colored shells in the middle panel of Figure \ref{shellvec_spect}, and compare with the 2D density plot in the top panel). We then extract electron spectra from individual shells (bottom panel in Figure \ref{shellvec_spect}, with the same color coding as in the middle panel; we only plot spectra for the shells belonging to the boundary island). The total spectrum obtained by integrating over the whole layer is shown with a solid black line. We see that, for every shell, the spectrum is distinctly non-thermal (compare with the Maxwellian plotted as a dashed black line), with a pronounced high-energy tail whose power-law slope is nearly the same in all the shells.

An earlier study of non-relativistic reconnection (with $\sigma$ ranging from 0.001 to 0.1, and $\beta$ from 0.02 to 0.2) argued that the power-law spectrum resulting from reconnection may not be a genuine power law, but it may rather result from the superposition of a series of Maxwellian distributions with spatially-varying temperatures \citep{li2017}. In contrast, we find that our spectra are genuine power-law distributions, at all locations inside the boundary island. The difference between our conclusions and the findings by \citet{li2017} might be attributed to the different regime of magnetization and plasma-$\beta$ that we explore.

\FloatBarrier

\bibliography{david_bib}
\bibliographystyle{apj}

\end{document}